\RequirePackage{fix-cm}
\documentclass[12pt]{article}
\RequirePackage{iftex}
\ifPDFTeX\pdfoutput=1\fi
\usepackage[utf8]{inputenc}
\usepackage[T1]{fontenc}
\usepackage{float}
\usepackage{authblk}
\usepackage[normalem]{ulem}
\usepackage{xcolor}
\usepackage{amsmath}
\usepackage{amsthm}
\usepackage{amssymb}
\usepackage{amsfonts}
\usepackage{algorithm}
\usepackage{algpseudocode}
\usepackage{adjustbox}
\usepackage{bbm}
\usepackage{graphicx}
\usepackage{booktabs}
\usepackage{mathtools}
\usepackage{multirow,array}
\usepackage{fullpage}
\usepackage{comment}
\usepackage[mathlines]{lineno}
\usepackage{subcaption}
\usepackage{bm}
\usepackage{siunitx}
\usepackage{microtype}
\ifPDFTeX
  \usepackage[style=numeric,sorting=none,maxnames=99]{biblatex}
\else
  \usepackage[backend=bibtex,style=numeric,sorting=none,maxnames=99]{biblatex}
\fi

\usepackage{tabularx,caption,longtable}

\usepackage[colorlinks = true,
            linkcolor = teal,
            urlcolor  = teal,
            citecolor = teal]{hyperref}

\DeclareRobustCommand{\statecode}[1]{%
\ifmmode
\text{\normalfont\ttfamily\detokenize{#1}}%
\else
{\normalfont\ttfamily\detokenize{#1}}%
\fi
}

\graphicspath{{./figures/}}

\newcommand{\papertitle}{Structural pressure networks and the collective resolution of high-pressing sequences in football}

\title{\papertitle}

\author[1,2]{Youcheng~Song}
\author[2]{Maddalena~Torricelli}
\author[2,3,4,5]{Brennan~Klein\thanks{\href{mailto:b.klein@northeastern.edu}{b.klein@northeastern.edu}}}

\affil[1]{Department of Statistics, University of Illinois Urbana-Champaign, Urbana, Illinois, USA}
\affil[2]{Network Science Institute, Northeastern University, Boston, Massachusetts, USA}
\affil[3]{Department of Communication Studies, Northeastern University, Boston, Massachusetts, USA}
\affil[4]{Department of Physics, Northeastern University, Boston, Massachusetts, USA}
\affil[5]{Institute for Experiential AI, Northeastern University, Boston, Massachusetts, USA}

\begin{document}
\maketitle
\pagenumbering{arabic}

\begin{abstract}
In football, whether a high press succeeds depends on more than just the defender closing down the ball carrier. Here, we represent each high-pressure event as a \textit{structural pressure network}, a graph-based representation that captures pressure on the carrier and possible receivers, passing lanes, player space, and overall team shape. We use features of these networks to predict whether a press ends in success, failure, or neither, fitting the model to events from 1.2 million pressing sequences in 9,220 league matches. At the onset of a pressing sequence, the model distinguishes successful presses from other outcomes. 
Averaging the difference between the model's predicted probabilities of success and failure over a team's presses in a season can define an expected press value---how promising a team's presses are, on average, when they begin---which is positively correlated with expected-goal difference per match in each of the 29 league-seasons. Lastly, this framework permits counterfactual positioning experiments in which we move one pressing player slightly and examine how a related model's prediction changes. More broadly, information about the collective arrangement of players (i.e., the network) can help us understand pressing across football.
\end{abstract}

\section{Introduction}\label{sec:intro}

In football, the team without the ball often tries to win it back by pressing, closing down the player in possession to force a turnover or at least to stop the attack from developing~\cite{Robberechts2019_pressing, Merckx2021_pressing}. When a team presses high, it tries to win the ball back near its opponent's goal, and whether a press succeeds depends on more than just the defender closing down the ball carrier~\cite{Merckx2021_pressing, Forcher2024_prediction}. It also depends on where that defender's teammates are, i.e.,~which receivers they mark, which passing lanes they close, and how much space they leave the opponent to play into~\cite{VanRoy2021_buildups, Lee2025_express, Forcher2024_prediction}. A teammate who cuts off a forward pass, for example, can force the carrier to play backward even when the pressing team does not regain possession. And while a fan or analyst watching the play may see how that teammate's positioning shaped the carrier's decision, it is less clear how to quantify that contribution when the forward pass was never attempted.

The simplest measures of pressing come from event data, which record when and where attacking actions (e.g., passes, carries, shots, and dribbles) and defensive actions (e.g., tackles, interceptions, and clearances) take place. Hudl StatsBomb, for instance, records ``pressure'' as a separate event when a defender closes down the player on the ball, identifying the defender and logging where the pressure began and how long it lasted~\cite{Morgan2018_counterpressing}. Event data of this kind can be used both to study trends in how teams play under pressure and to develop metrics and models of how they press their opponents. A recent study of ten top-flight men's and women's competitions found that the number of passes completed under pressure per match rose between 2020 and 2025~\cite{Carstens2026_football}. From the defending team's perspective, a common measure of pressing intensity is passes per defensive action, which counts how many passes a team allows its opponent in a given area of the pitch for each defensive action it takes there~\cite{Bekkers2024_pressing}. The VPEP model (Valuing Pressure decisions by Estimating Probabilities) goes a step further, using recorded pressure events to estimate how a decision to press changes the team's chances of recovering the ball and its risk of conceding a shot~\cite{Robberechts2019_pressing}.

To understand how the players supporting a press constrain or enable the options available to the ball carrier, we also need to know where those players are as the press unfolds. Tracking data record every player's position several times a second, making it possible to estimate how much pressure a defender puts on a potential receiver before the carrier attempts a pass~\cite{Andrienko2017_pressure, Merckx2021_pressing}. Across the pitch, these estimates can describe how intensely each defender presses each attacker~\cite{Bekkers2024_pressing} or, for example, how pressure is distributed across the team in possession~\cite{Gu2024_pressuremap}. Tracking data have also been used to predict whether the pressing team will regain the ball~\cite{Forcher2024_prediction} and to examine how the likelihood of regaining the ball changes when one defender is moved to a different position in the model while the other players are held fixed~\cite{Lee2026_express}.

Broadcast frames offer a middle ground between event data and tracking data. In its StatsBomb 360 data, Hudl uses a shared event ID to link a snapshot from the television broadcast to the corresponding event record. This frame then provides the position and team of each player in the camera's view at that moment, which can be combined with the event record that identifies who performed the action, what they did, and other associated event metadata~\cite{Statsbomb2025_data}. Researchers have used StatsBomb 360 data to study how teams disrupt an opponent's build-up~\cite{VanRoy2021_buildups}, evaluate team defense from the positions of the visible players~\cite{Umemoto2025_valuation}, estimate individual players' contributions to a press~\cite{Lee2025_express}, and compare defenders' observed positions with counterfactual alternatives~\cite{Umemoto2023_counterfactuals}. Because a frame shows several players at the same moment, we can also represent those players as nodes in a \textit{network} and use links to describe the relationships among them. In football, networks have been used to characterize teams' passing structures~\cite{Buldu2018_passing, Grund2012_network, Buldu2019_guardiola} and to describe team shape in terms of players' spatial connectivity on the pitch~\cite{Brandes2025_shape}. Researchers have also built networks directly from broadcast frames to study when teams concede goals~\cite{Zhang2026_bonds} and which teammate receives a pass~\cite{Wang2026_receiver}.

Here, we ask what the network configuration of the two teams can tell us about how a pressing sequence will end. To describe that configuration, we represent the players visible in a broadcast frame as nodes in a \textit{structural pressure network}. We use two types of links to connect players in this network: 1) \textit{passing links}, which trace the lanes from the ball carrier to teammates who could receive a pass, and 2) \textit{pressure links}, which describe how much pressure the defending players apply to the carrier and potential receivers. We also measure how much space is available to each player and describe each team's shape by identifying which teammates are adjacent on the pitch and how far apart they are~\cite{Brandes2025_shape}. To test how much this collective information adds to our ability to predict the outcome of a press, we compare a model based on the carrier's position, recent play, and direct pressure on the ball with one that also accounts for the arrangement of the surrounding players. We then use the full model to measure how promising a team's presses are, on average, when they begin, and examine how this \textit{expected press value} relates to expected-goal difference per match. Finally, we use a similar model to estimate how the chance of regaining the ball would change if the player closing down the carrier were in a slightly different position when they applied pressure. For each press, we compare the predicted regain probability at the player's true observed position with the mean across nearby alternatives; the average difference across a player's presses can then serve as a proxy for the effectiveness of their positioning. In sum, the network lets us evaluate pressing as a collective effort, using the arrangement of both teams to better predict how a press will end and assess how well the player closing down the ball is positioned.

\section{Results}\label{sec:results}

\subsection{The press as a network}\label{sec:results_spn}

\begin{figure}[t]
    \centering
    \includegraphics[width=0.95\linewidth]{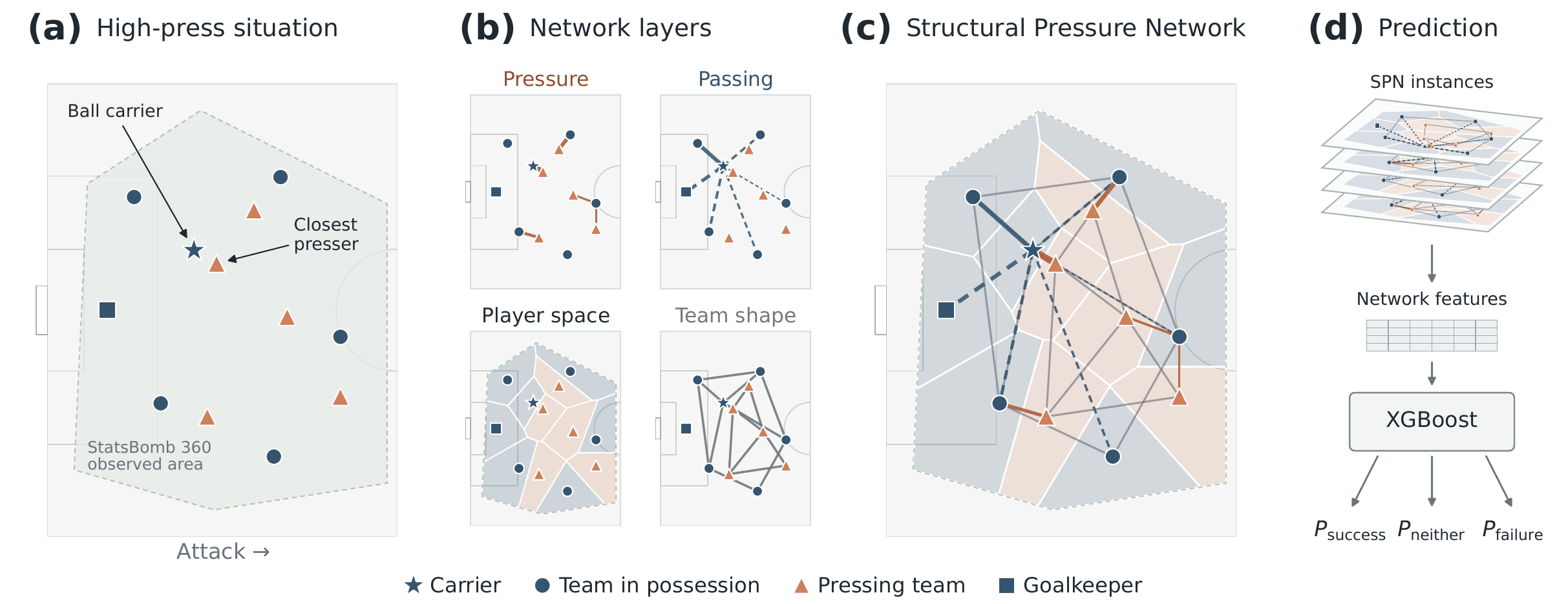}
    \caption{\textbf{Constructing and using structural pressure networks.} \textbf{(a)} Schematic of a high-press situation within the StatsBomb 360 broadcast frame observed area. \textbf{(b)} Four representations of the same situation: pressure links, passing options, player space (Voronoi cells), and team shape graph~\cite{Brandes2025_shape}. \textbf{(c)} These layers combine into a structural pressure network. \textbf{(d)} Features extracted from individual networks enter an XGBoost classifier, which estimates the probabilities that a pressing sequence will end in success, neither, or failure.}
    \label{fig:spn_overview}
\end{figure}

We construct a structural pressure network for each high-pressure event using the positions of the ball carrier and the other players visible in the associated StatsBomb 360 frame (Fig.~\ref{fig:spn_overview}a). We call a pass, carry, or dribble a \textit{high-pressure event} when a player makes it in their team's own half under close pressure from an opponent. A \textit{pressing sequence} consists of one or more such events during a continuous spell of pressure on the same team, with at most five seconds between the end of one event and the start of the next (Section~\ref{sec:sequence}).

The network combines four representations of the same spatial configuration (Fig.~\ref{fig:spn_overview}b,c). (i) Pressure links connect pressing outfield players to the ball carrier and possible receivers, with larger weights for closer opponents. (ii) Passing links connect the carrier to possible receivers, with weights that decrease with passing distance and obstruction by pressing players. (iii) Voronoi cells describe the space closest to each player within the observed area. (iv) A \textit{shape graph} for each team connects neighboring teammates and describes the spacing between them~\cite{Brandes2025_shape}. Together, these representations describe how pressure on the ball relates to the passing options and open space around it.

We extract features from each network, including the total pressure on the carrier, the number of open passing options, how evenly pressure is spread across the carrier and possible receivers, and how much space is available to players ahead of the ball. For each high-pressure event, we calculate 26 model inputs. Nine describe the carrier's location, recent play, and direct pressure on the ball. The other 17 describe passing options, the distribution of pressure across the carrier and possible receivers, player space, and team shape. Two presses can put similar pressure on the carrier, for example, while only one leaves an open forward pass to a teammate under little pressure. We use these inputs to fit a single gradient-boosted classifier (XGBoost) across five leagues (Fig.~\ref{fig:spn_overview}d). Each event supplies one training example, labeled with the eventual outcome of its pressing sequence.

The classifier estimates the probabilities that a pressing sequence ends in success, failure, or neither. A sequence ends in \textit{success} when the pressing team regains possession or wins a favorable stoppage. \textit{Failure} includes the opponent escaping the press or the pressing team committing a foul. \textit{Neither} describes a resolved sequence in which the opponent retains possession without escaping, such as a return to its goalkeeper or a retained throw-in. We excluded sequences whose outcome could not be determined. We evaluate predictions at sequence onset, using the first high-pressure event to predict the sequence's eventual outcome.

\subsection{Predicting the resolution of pressing sequences}\label{sec:results_prediction}

\begin{figure}[t!]
    \centering
    \includegraphics[width=0.95\linewidth]{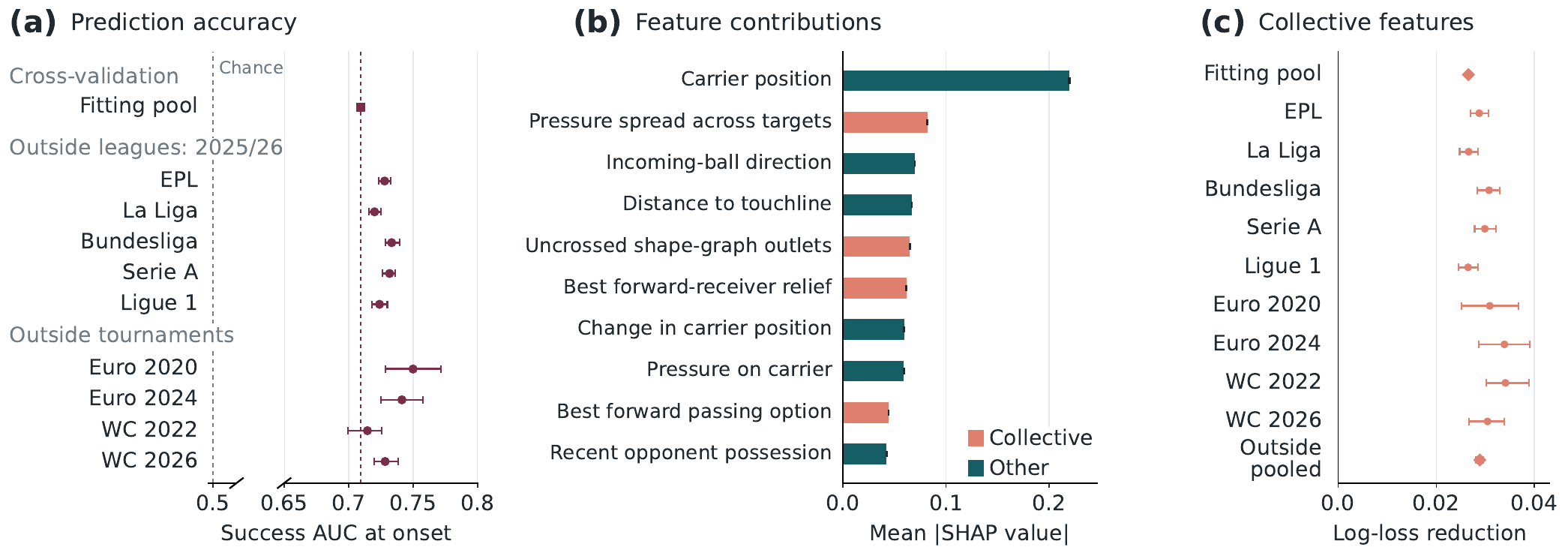}
    \caption{\textbf{Predicting the resolution of pressing sequences.} \textbf{(a)} Success AUC at sequence onset in cross-validation and nine outside datasets. The dashed line marks cross-validated performance. \textbf{(b)} The ten features with the largest mean absolute TreeSHAP contributions at sequence onset in the nine outside datasets, colored by feature group. \textbf{(c)} Reduction in onset log loss when collective network features are added to the matched comparison model. Positive values indicate that the collective network features improve the model. Error bars show 95\% match-bootstrap intervals.}
    \label{fig:prediction}
\end{figure}

At the onset of a pressing sequence, the model distinguishes sequences that go on to succeed from those ending in failure or neither (Fig.~\ref{fig:prediction}a). We fitted the classifier to events from 1,245,248 sequences across 9,220 matches in 24 men's league-seasons of the Premier League, La Liga, Bundesliga, Serie A, and Major League Soccer (MLS). In five-fold cross-validation, keeping whole matches together, the area under the receiver operating characteristic curve (AUC) for success at onset was 0.709 (95\% match-bootstrap interval $0.708$--$0.710$). We then applied the fitted model without refitting to 314,030 sequences from 2,006 matches in five outside league-seasons and four international tournaments. AUCs ranged from 0.720 to 0.733 in the five league-seasons and from 0.714 in the 2022 World Cup to 0.750 in Euro 2020. Across the five outside league-seasons, presses succeeded more often and failed less often than the model predicted, so the predicted difference between the success and failure rates was 8.7 percentage points below the observed difference (Supplementary Section~\ref{app:calibration}).

To examine which inputs contributed most to the model's predictions, we used TreeSHAP~\cite{Lundberg2020_treeshap} to estimate each input's contribution to the three outcome scores at sequence onset. For each input, we averaged the absolute contributions across outcomes and all 314,030 onsets in the nine outside datasets (Fig.~\ref{fig:prediction}b). The carrier's position along the pitch contributed most on average, followed by how evenly pressure was distributed across the carrier and possible receivers and then by the direction from which the ball had arrived. Four of the ten leading inputs were collective features, including the number of passing lanes to outfield teammates that crossed no edge of the pressing team's shape graph and a measure that combined the openness of a forward passing lane with how little pressure its receiver faced.

To test how much information about the surrounding players added to prediction, we compared the full model with a model fitted to the same events using only the nine inputs describing the carrier's location, recent play, and direct pressure on the ball. We compared their predictions at onset using log loss, which penalizes a model more heavily if it assigns low probability to an outcome that actually follows. Adding the 17 collective features reduced log loss from 0.970 to 0.943 in cross-validation and from 0.996 to 0.967 across the pooled outside datasets, relative reductions of 2.7\% and 2.9\%, respectively (Fig.~\ref{fig:prediction}c). The full model improved predictions in each of the nine outside datasets, with all nine paired 95\% match-bootstrap intervals for the reduction in log loss above zero.

\subsection{Expected press value and team performance}\label{sec:results_fingerprints}

\begin{figure}[t!]
    \centering
    \includegraphics[width=0.95\linewidth]{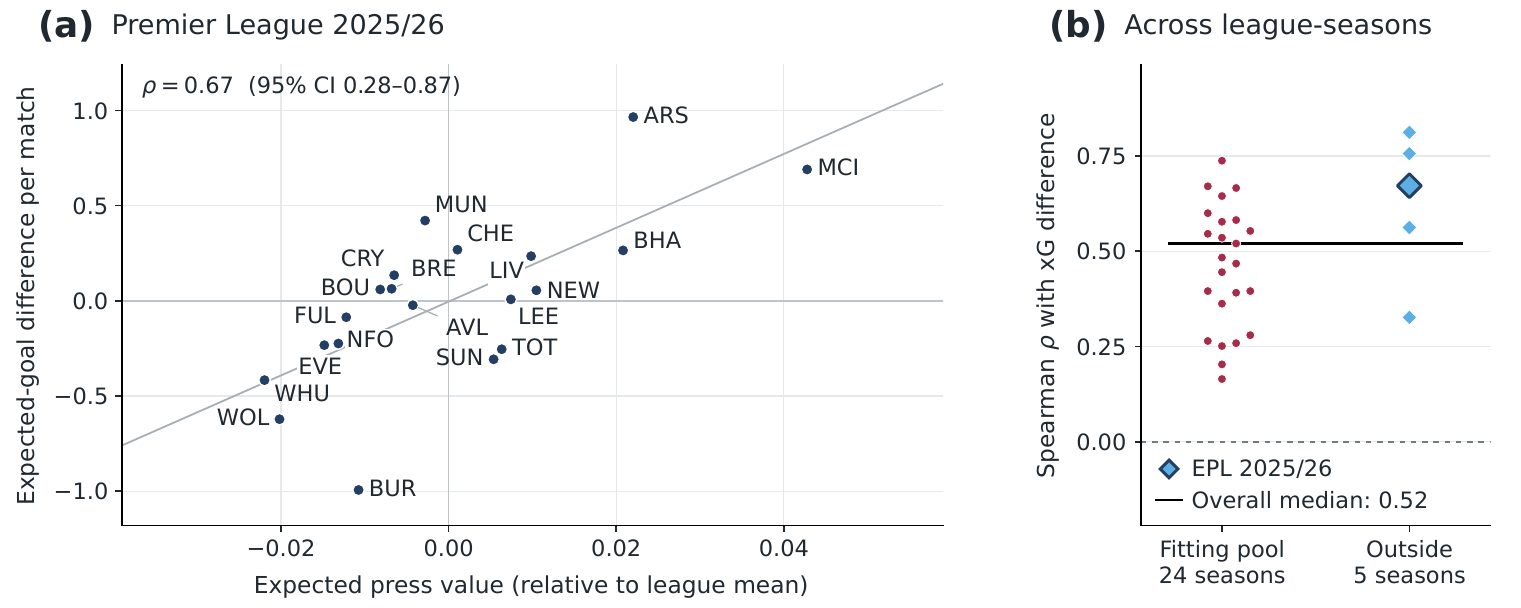}
    \caption{\textbf{Expected press value and team performance.} \textbf{(a)} Expected press value, centered on the league-season mean, versus expected-goal difference per match for Premier League teams in 2025/26. The line shows a least-squares fit, and the annotation gives Spearman's $\rho$ and its approximate 95\% interval. \textbf{(b)} Within-season correlations across 29 league-seasons. Each point represents one league-season. Crimson circles identify fitting-pool seasons and light-blue diamonds identify outside seasons. The outlined diamond highlights Premier League 2025/26, and the horizontal line marks the overall median.}
    \label{fig:team_value}
\end{figure}

We use the model's predictions to summarize how promising a team's presses are, on average, when they begin. For each pressing sequence, we subtract the predicted probability of failure from the predicted probability of success at onset. Averaging these differences over team $t$'s eligible presses in a league-season gives its \textit{expected press value},
\begin{equation}
	V_t = \frac{1}{N_t}\sum_{s \in \mathcal{S}_t}\left(p_s^{+}-p_s^{-}\right),\label{eq:team_expected_press_value}
\end{equation}
where $\mathcal{S}_t$ contains the team's $N_t$ eligible pressing sequences, and $p_s^{+}$ and $p_s^{-}$ are the predicted probabilities of success and failure at the onset of sequence $s$. Each press contributes equally to the team's average. For Fig.~\ref{fig:team_value}a, we center each team's expected press value by subtracting the mean success-minus-failure prediction across all eligible presses in its league-season. A team to the right of zero therefore began its presses in situations the model rated more favorably than the league-season average.

To assess how expected press value relates to team performance, we compared it with expected-goal difference per match within each league-season. We calculated expected-goal difference per match by subtracting a team's expected goals conceded from its expected goals created and dividing by matches played. Then for the 24 fitting-pool seasons, we scored each press with a classifier trained on other matches. For the five outside seasons, we used the classifier trained on all 24 seasons. In the Premier League in 2025/26, Spearman's rank correlation was $\rho=0.67$ (approximate 95\% interval $0.28$--$0.87$, Fig.~\ref{fig:team_value}a). The correlation was positive in all 29 league-seasons (median $\rho=0.52$, range $0.16$--$0.81$, Fig.~\ref{fig:team_value}b). Approximate 95\% intervals excluded zero in 17 seasons.

Within a league-season, teams with higher expected-goal difference tended to begin their presses in situations the model rated more favorably. Complete season estimates and comparisons with points per match are reported in Supplementary Section~\ref{app:teams}.

\subsection{Counterfactual positioning experiments}\label{sec:results_mechanism}

\begin{figure}[t!]
    \centering
    \includegraphics[width=0.95\linewidth]{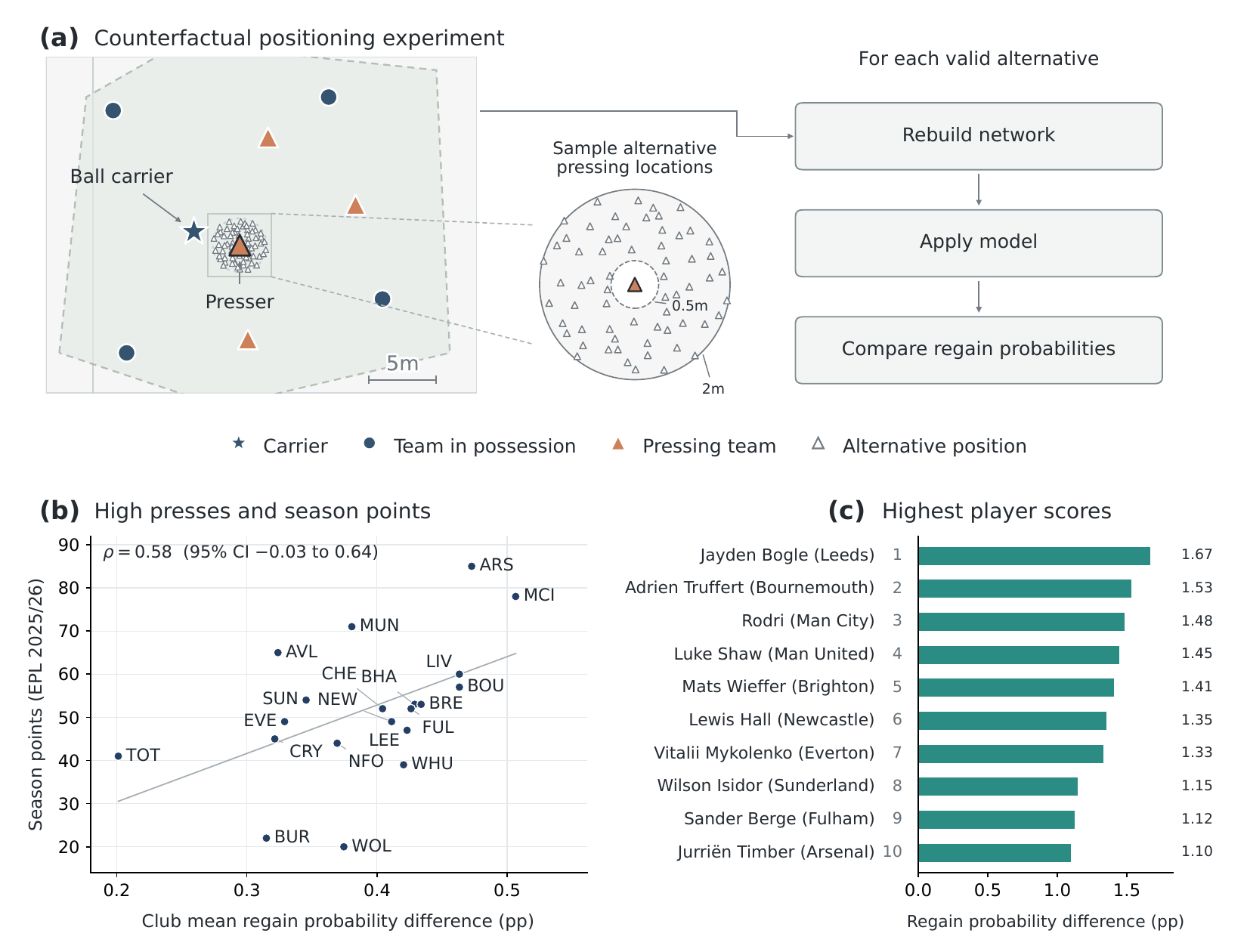}
    \caption{\textbf{Model-based position quality during Premier League high presses.} \textbf{(a)} Schematic of the counterfactual positioning experiment. We move the recorded presser to nearby positions while holding the other players fixed. At each valid alternative, we rebuild the structural pressure network and predict the probability of a controlled regain within five seconds. Position quality is the predicted regain probability at the observed position minus the mean across valid alternatives, expressed in percentage points. \textbf{(b)} Correlation between club mean position quality and the 2025/26 season points for all 20 Premier League clubs. \textbf{(c)} The ten highest player means among the 181 players with at least 15 eligible high presses and at least 1,350 Premier League minutes (15 full-match equivalents) that season.}
    \label{fig:counterfactual_positions}
\end{figure}

We next asked how the predicted chance of regaining the ball would change if the defender closing down the carrier had applied pressure from a slightly different position (Fig.~\ref{fig:counterfactual_positions}a). For this analysis, we used recorded StatsBomb pressure events, which identify the defender applying pressure. We fitted a separate classifier to 20,293 eligible pressure events in the 2025/26 Premier League to predict whether the pressing team would regain controlled possession in play within five seconds. We used the 8,607 events that met our high-press criteria for the positioning experiment.

For each high press, we sampled 64 candidate positions between 0.5 and 2 meters from the presser's observed location (Fig.~\ref{fig:counterfactual_positions}a). At each valid alternative, we moved the presser, held the other players and the preceding play fixed, and rebuilt the structural pressure network. We scored the observed position and its alternatives with the same classifier, fitted with that match held out. We define \textit{model-based position quality} as the predicted regain probability at the observed position minus the mean across its valid alternatives, expressed in percentage points. A positive value means that, given how the play had developed, the model rated the presser's observed position more favorably than the nearby alternatives on average.

Across the 8,607 eligible high presses, the pressing team regained controlled possession within five seconds in 17.1\% of cases. The predicted regain probability at the presser's observed position exceeded the mean across its nearby alternatives in 55.2\% of presses, with an average difference of 0.39 percentage points across all presses. Giving each eligible press equal weight, mean position quality was positive for all 20 clubs, ranging from 0.20 to 0.51 percentage points. Clubs with higher mean position quality tended to finish the season with more points (Spearman's $\rho=0.58$, 95\% fixture-bootstrap interval $[-0.03,\,0.64]$, Fig.~\ref{fig:counterfactual_positions}b).

Among the 181 players with at least 15 eligible high presses and at least 1,350 Premier League minutes, Jayden Bogle had the highest mean position quality (Fig.~\ref{fig:counterfactual_positions}c). Across his 22 eligible presses, the model assigned his observed positions a regain probability 1.67 percentage points higher than that of his nearby alternatives on average (95\% fixture-bootstrap interval $[0.87,\,2.60]$). Adrien Truffert followed at 1.53 percentage points across 18 presses and Rodri at 1.48 across 16. See Supplementary Table~\ref{tab:si-high-press-position-quality} for full player results.

\section{Discussion}\label{sec:discussion}

We developed a model that estimates, at the start of a high press, whether the pressing sequence will end in success, failure, or neither, building on earlier work that used player positions to measure pressure on potential receivers~\cite{Merckx2021_pressing} and predict whether the pressing team would regain the ball~\cite{Forcher2024_prediction}. Fitted to 1.2 million pressing sequences from five top-flight men's leagues, our model distinguished successful presses from other outcomes in cross-validation (AUC of 0.709) and when applied without refitting to five outside league-seasons and four international tournaments (AUCs from 0.714 to 0.750). Information about the network arrangement of the surrounding players, e.g.,~passing options the carrier never used, improved these predictions beyond the carrier's location, recent play, and direct pressure on the ball. Adding these collective features from our structural pressure networks reduced onset log loss in each of the nine outside datasets, with a 2.9\% reduction across the pooled outside data---evidence that the collective arrangement of off-ball players helps predict how a press will end. While it may seem obvious that off-ball positioning matters to a press, representing both teams in a network gives us a way to begin quantifying what the surrounding players contribute beyond the defender closing down the ball.

At the start of each pressing sequence, our model estimates how likely the press is to succeed and how likely it is to fail. Averaging the difference between these two probabilities over all of a team's pressing sequences across a given season gives its expected press value, a measure we introduced to quantify how promising the team's presses are when they begin. This complements measures of pressing intensity, such as passes per defensive action~\cite{Bekkers2024_pressing}, because a team can press infrequently and still choose favorable moments to do so. We found that teams with higher expected press value also tended to have a higher expected-goal difference per match, which is encouraging for a measure calculated before the outcome of each press is known. Teams with higher expected-goal difference also tended to begin their presses closer to the opponent's goal (Supplementary Section~\ref{app:teams}). Expected press value reflects where a team's presses begin on the pitch as well as the arrangement of players around the ball. Because the measure captures both, it lets a coaching staff compare presses that begin in similar areas on the pitch but are set up differently, then estimate how a novel player network arrangement could change the chance of success.

The network can also be used to evaluate the positioning of the player closing down the ball. We fitted an additional model that predicts the probability that the pressing team will regain possession within five seconds of a press. Following earlier studies that moved a defender and recalculated a model's output~\cite{Umemoto2023_counterfactuals, Lee2026_express}, we used the predicted chance of a regain at the presser's \textit{observed} position as a baseline and compared it with the average across alternate positions 0.5--2 meters away. These small changes in the presser's location can qualitatively change the situation, for instance by adding pressure on the ball or opening a passing lane the presser had been blocking. As such, we rebuilt the structural pressure network at each alternative position, ran each counterfactual network through the model, and recorded the predicted regain probability. Across the eligible Premier League high presses, 17.1\% of which ended in a controlled regain, players' observed positions had a predicted regain probability 0.39 percentage points higher, on average, than the nearby alternatives. If a player's observed positions score above the nearby alternatives on average, this may reflect strong positional awareness; averaging this difference over each player's presses gives a ranking of players by the model's assessment of their positioning, an interpretable measure that analysts could use to identify positionally sound players for recruitment or to work on positioning in training.

\begin{table}[t]
    \centering
    \caption{\textbf{Pressing efficiency and position quality at the 2026 men's World Cup.} Pressing efficiency is the mean first-to-last change in $P(\mathrm{success})-P(\mathrm{failure})$ across linked sequences, using a model fitted to the tournament. Position quality compares observed and nearby alternative positions using the five 2025/26 Premier League regain models, applied without refitting.}
    \label{tab:world_cup_players}
    \small
    \setlength{\tabcolsep}{4pt}
    \begin{adjustbox}{max width=\linewidth}
    \begin{tabular}{@{}r lrr @{\hspace{1.5em}} lrr@{}}
        \toprule
        & \multicolumn{3}{c}{\textbf{Pressing efficiency}} & \multicolumn{3}{c}{\textbf{Position quality}} \\
        \cmidrule(lr){2-4}\cmidrule(lr){5-7}
        Rank & Player (team) & Sequences & Score (pp) & Player (team) & High presses & Score (pp) \\
        \midrule
         1 & Elliot Anderson (ENG)  & 6 & $+41.36$ & Moisés Caicedo (ECU)    &  5 & $+1.86$ \\
         2 & Agustín Canobbio (URU) & 7 & $+40.53$ & Julián Álvarez (ARG)    &  6 & $+1.53$ \\
         3 & Amine Gouiri (ALG)     & 5 & $+38.82$ & Dayot Upamecano (FRA)   &  5 & $+1.40$ \\
         4 & Federico Viñas (URU)   & 5 & $+32.06$ & Rodri (ESP)             &  6 & $+1.06$ \\
         5 & Mikel Oyarzabal (ESP)  & 6 & $+27.66$ & Nathaniel Brown (GER)   &  5 & $+1.04$ \\
         6 & Farès Chaïbi (ALG)     & 6 & $+26.81$ & Pedro Vite (ECU)        & 10 & $+0.89$ \\
         7 & Romano Schmid (AUT)    & 6 & $+25.19$ & Azzedine Ounahi (MAR)   &  6 & $+0.88$ \\
         8 & Nathaniel Brown (GER)  & 5 & $+20.10$ & Breel Embolo (SUI)      &  9 & $+0.87$ \\
         9 & Azzedine Ounahi (MAR)  & 5 & $+17.88$ & Farès Chaïbi (ALG)      &  8 & $+0.79$ \\
        10 & Michael Olise (FRA)    & 7 & $+17.67$ & Eldor Shomurodov (UZB) &  5 & $+0.74$ \\
        \bottomrule
    \end{tabular}
    \end{adjustbox}
\end{table}

More topically, we used the 2026 men's World Cup to examine pressing efficiency and individual positioning in a tournament setting (Table~\ref{tab:world_cup_players}), finding qualitatively reasonable player rankings. The pressing-efficiency ranking is a player-level analogue of the team comparison in Fig.~\ref{fig:team_value}a; for teams, we averaged the difference between the predicted probabilities of success and failure at the start of their presses. For players, we used a sequence model fitted to the tournament and averaged the change in that difference from the first event to the last across the presses linked to each player (Supplementary Section~\ref{app:world_cup_efficiency}). England's Elliot Anderson ranked first among players with at least five linked sequences and at least 180 tournament minutes. Across the six pressing sequences linked to him, England's predicted chance of success rose by 19.12 percentage points on average from the first event to the last, while its predicted chance of failure fell by 22.24 points. For positioning, we repeated the counterfactual player comparison in Fig.~\ref{fig:counterfactual_positions}c, applying our regain model to the tournament without refitting. We compared the predicted chance of regaining the ball within five seconds at each presser's observed position with the average across nearby alternatives, and by this measure, Moisés Caicedo ranked first among the players with at least five high presses and who played at least 180 tournament minutes, followed by Julián Álvarez. Together, these World Cup rankings illustrate two ways to use the arrangement of both teams to evaluate a player's role in a collective press. Across several league seasons, we could use the same approach to examine how a player's pressing varies against different opponents, with enough matches to revisit situations that arise only a few times in a single tournament.

Three limitations affect how these results should be used and suggest where the model could be improved and extended. First, how closely the predicted probabilities match observed outcomes varies across competitions: across the five outside league-seasons, the predicted difference between the success and failure rates was 8.7 percentage points below the observed difference, so expected press value is best compared between teams in the same league-season (Supplementary Section~\ref{app:calibration}). Second, half a season provides too few high presses to separate lasting positional habits from the particular situations a player faced; rankings before and after January 1 agreed weakly (Spearman's $\rho=0.07$), but random splits of each player's own matches yielded a median correlation of only 0.13 (Supplementary Section~\ref{app:positioning_checks}), so testing persistence will require pooling multiple seasons. Third, our analyses cover only men's football. Since 2020, for example, the spatial structure and passing networks have changed more in elite women's football than in men's on many measures~\cite{Carstens2026_football}, and in general, we may not expect insights from models trained on men's competitions to carry over directly to women's~\cite{Rubenstein2026_nbawnba, Minuci2026_wnba}. Repeating the comparison of feature sets in women's competitions would show how much the collective features add in that setting.

Moving forward, our experiments can be extended to the off-ball teammates who support a press away from the ball. A broadcast frame shows where such a teammate stood if they were in the camera's view, but the event record names only the presser, so we evaluated the presser's position given where the teammates stood and left the contribution of any named teammate unmeasured. With the presser held in place, we could move the teammate covering a forward receiver and estimate how the predicted chance of a regain changes, as has been illustrated with tracking data~\cite{Lee2026_express}. Tracking data would let us identify the teammate across presses, follow how receivers moved as defenders closed on them, and see what triggers teammates to cover certain passing lanes.

Which outcome we ask the model to predict also matters; for example, a high press that forces an opponent to play back to the goalkeeper may have done exactly what the pressing team intended. In our sequence analyses, a press that ended with the opponent retaining possession through a pass back to its goalkeeper counted as \textit{neither}. The same networks could be used to study which arrangements force play backward or prevent an opponent from playing through the first line of pressure~\cite{VanRoy2021_buildups}. Ultimately, we hope this work encourages others to try different predictors and to choose outcomes that reflect what a team is trying to achieve. In each case, however, we argue that the network will continue to be an informative mathematical syntax for capturing the complexity of football positional tactics, allowing us to simultaneously value the player closing down the ball and the teammates covering the next pass in the same model.

\section{Methods \& Data}\label{sec:methods}

\subsection{Data and preprocessing}\label{sec:data}

We used licensed Hudl StatsBomb event and 360 data from men's competitions~\cite{Statsbomb2025_data}, where event data records each action and the player who performed it, and the corresponding 360 frame gives the positions and teams of the players in the broadcast frame. We constructed each network from the carrier's location and the positions of the surrounding players visible in that frame.

We fitted the sequence classifier using matches from the Premier League, La Liga, Bundesliga, and Serie A from 2020/21 through 2024/25 and Major League Soccer from 2021 through 2024. These 24 league-seasons contributed 2,562,468 high-pressure events in 1,245,248 resolved pressing sequences from 9,220 matches. We evaluated the classifier without refitting on nine outside datasets: the available 2025/26 matches from the four European fitting leagues and Ligue 1, Euro 2020, Euro 2024, and the 2022 and 2026 World Cups. Together they contributed 314,030 resolved sequences from 2,006 matches (Supplementary Table~\ref{tab:population_flow}).

To recover the positions of the players at the time of a given action, we matched its event record to the corresponding 360 frame by identifier. For approximately one third of the fitting matches, the provider data came with frame identifiers that did not have a corresponding event identifier, so we needed to use event order and the location of the player performing the action to link the two records (Supplementary Section~\ref{app:frame_linkage}). Note that none of the 2,006 outside matches required this extra step.

For the counterfactual positioning experiments, we sampled alternative positions between 0.5 and 2 meters from the presser's observed position. To measure those distances in meters, we rescaled the provider's $120\times80$ coordinate grid to a pitch measuring $105\times68$ meters. We used this scale throughout and oriented the pitch so that increasing $x$ always meant moving toward the opponent's goal from the perspective of the team in possession.

\subsection{High-pressure events and sequence outcomes}\label{sec:sequence} 

To identify occasions when a team came under pressure in its own half, we considered its passes, carries, and dribbles, excluding play patterns originating from free kicks, corners, throw-ins, and kick-offs. At each candidate event, we calculated a pressure score from the distances between the carrier and the visible opposing outfield players. The score also accounts for the nearest pitch boundary, since it constrains the space into which the carrier can move. We classified a candidate action as a high-pressure event if its pressure score exceeded 0.65 and at least one opposing outfield player was within 6.4 meters of the carrier (Supplementary Section~\ref{app:detector_details}).

A press can continue as the ball gets passed between teammates. We grouped successive high-pressure events for the same team in possession within the same period, allowing at most five seconds between the end of one event and the start of the next. An intervening on-ball action by the opponent, an observed release from pressure, an escape across halfway, or a stoppage would end the sequence. A change in the provider's possession identifier alone did not end the sequence, since the possession identifier could change even between a player's carry and the pass they made immediately afterward. And finally, we refer to the first qualifying event as the sequence's \textit{onset}.

Starting at the final high-pressure event, we followed play until we could determine the outcome or the period ended. A regain or favorable stoppage counted as \textit{success} for the pressing team. An escape by the opponent or a foul by the pressing team counted as \textit{failure}. If the opponent retained possession without escaping, the outcome was \textit{neither}. We excluded unresolved sequences. We also required an identifiable carrier and a network with at least four players at every high-pressure event, so a sequence with a usable onset could still be excluded if a later event could not be represented (Supplementary Sections~\ref{app:sequence_resolution} and~\ref{app:cohort_flow}).

\subsection{Structural pressure networks and features}\label{sec:spn}

For each high-pressure event, we used the visible players as nodes and constructed two types of links (Fig.~\ref{fig:spn_overview}). Pressure links connect an opposing outfield player to the carrier or a possible receiver, and their weight increases as the opponent gets closer. Passing links connect the carrier to teammates who could receive the ball, and their weight decreases as the receiver gets farther away and as an opposing outfielder stands closer to the line of the pass. We measured the space available to each player using a Voronoi tessellation of the visible players' locations (clipped to the part of the pitch visible in the frame). Each player's cell is the area that is closer to that player than to any other visible player. To describe team shape, we used the \textit{shape graph} of Brandes et al.~\cite{Brandes2025_shape}, which connects each outfield player to neighboring teammates. Four inputs to our model use the lengths of these links, whether a passing lane crosses them, or whether an opponent stands on them.

Our sequence classifier used 26 inputs. Nine described the carrier's position, recent play, and direct pressure on the ball. Direct pressure combined the contributions of the visible opposing outfielders and, unlike the detection score in Section~\ref{sec:sequence}, excluded the pitch boundary. The other 17 inputs described passing options, how pressure was spread across the carrier and possible receivers, player space, and team shape. For example, one input combined each forward passing link's weight with the pressure on its receiver and kept the best option, measuring how readily the carrier could reach a forward teammate under little pressure. Inputs measuring change used differences from the preceding observation, without division by elapsed time. Supplementary Section~\ref{app:network} defines all 26 inputs, including the preceding observation used for each change and the handling of missing values.

\subsection{Model fitting and onset evaluation}\label{sec:methods_model}

We fitted a three-outcome XGBoost classifier~\cite{Chen2016_xgboost} across the five leagues. Each high-pressure event supplied a training example labeled with its sequence's eventual outcome. To give long and short presses equal total weight, we assigned each event a weight of $1/n$, where $n$ is the number of events in its sequence. We used XGBoost's handling of missing inputs and applied no additional weights to balance the outcomes. We selected the tree settings on a validation subset of the fitting pool and retained them across evaluation fits (Supplementary Section~\ref{app:model}).

We evaluated one prediction per sequence, at the onset of its first high-pressure event. In five-fold cross-validation, we kept whole matches together and predicted each fold with a classifier fitted to the other four. We calibrated the onset predictions by temperature scaling~\cite{Guo2017_calibration}, which divides the classifier's three outcome scores by a common positive value before converting them to probabilities. We estimated this value from predictions and observed outcomes in half the matches of each held-out fold and applied it to the other half, then exchanged their roles. Each evaluated match was therefore held out from both the classifier fit and the temperature estimate used to score it.

For the nine outside datasets, we fitted the classifier to all 24 fitting-pool seasons. To estimate a single calibration temperature, we used the onset predictions made when each fitting-pool match was held out and the observed outcomes of those sequences. We then applied the fitted classifier and temperature unchanged to all nine outside datasets. We used the AUC to measure how well the model distinguished successful presses from those ending in \textit{failure} or \textit{neither} and assessed calibration by comparing predicted probabilities with observed outcome frequencies. Supplementary Sections~\ref{app:model} and~\ref{app:calibration} give the interval procedures and the calibration comparison.

\subsection{Collective-feature comparison and feature contributions}\label{sec:methods_comparison}

To identify which inputs contributed most to the predictions, we calculated exact TreeSHAP contributions~\cite{Lundberg2020_treeshap} for the classifier fitted to the full fitting pool at all 314,030 onsets in the nine outside datasets (Fig.~\ref{fig:prediction}b). At each onset, TreeSHAP assigns each input a signed contribution to each of the three outcome scores. We divided the contributions by the fitted temperature, which puts them on the scale of the calibrated scores that are converted to the reported probabilities. We ranked inputs by their mean absolute contribution, giving equal weight to each onset and each outcome, so that contributions in opposite directions could not cancel. Supplementary Section~\ref{app:shap} gives the complete ranking.

To test whether the 17 collective features improved prediction, we compared the full classifier with a nine-input classifier that used only the inputs for the carrier's position, recent play, and direct pressure on the ball (Fig.~\ref{fig:prediction}c, Supplementary Table~\ref{tab:features}). We fitted both classifiers to the same events with the same outcomes, sequence weights, match folds, and tree settings, and estimated a separate temperature for each using the procedure in Section~\ref{sec:methods_model}.

We compared the classifiers using multiclass log loss, the mean negative logarithm of the probability assigned to the outcome that occurred~\cite{Gneiting2007_scoring}. We subtracted the full classifier's loss from the nine-input classifier's loss, so a positive difference indicates that adding the collective features improved prediction. We calculated this difference in cross-validation, in each outside dataset, and across all outside onsets pooled, giving each onset equal weight. For 95\% intervals, we drew whole matches with replacement 200 times and evaluated both classifiers on the same matches in each draw, re-estimating their temperatures and holding their fitted trees fixed (Supplementary Section~\ref{app:collective}).

\subsection{Expected press value and team performance}\label{sec:methods_value}

At the onset of each of a team's eligible presses, we subtracted the calibrated probability that the press would fail from the calibrated probability that it would succeed. We averaged these differences over the league-season, giving each press equal weight, to obtain the team's expected press value (Eq.~\ref{eq:team_expected_press_value}). For the 24 fitting-pool league-seasons, we used predictions from matches held out during fitting and calibration. For the five outside league-seasons, we used the classifier and temperature estimated from the full fitting pool. For Fig.~\ref{fig:team_value}a, we centered each team's value on the mean across all eligible presses in its league-season.

To calculate each team's expected-goal difference per match, we summed the StatsBomb expected-goal values of its own shots, subtracted the total for its opponents' shots, and divided by the number of matches it played. These values come from the event records, so the calculation included fixtures for which 360 data were unavailable. We used all available fixtures with event data and a recorded score, including MLS playoff fixtures, and excluded shoot-out attempts. Within each league-season, we calculated Spearman's rank correlation across teams between expected press value and expected-goal difference per match, with approximate 95\% intervals using the Bonett-Wright adjustment~\cite{Bonett2000_correlations}. Supplementary Section~\ref{app:teams} gives the correlation for each league-season, the corresponding correlations with points per match, and the analysis of where teams began their presses.

\subsection{Counterfactual positioning experiments}\label{sec:methods_positioning}

We compared the recorded presser's position with nearby alternatives, drawing on previous experiments that moved a defender and recalculated the predicted outcome~\cite{Umemoto2023_counterfactuals, Lee2026_express}. Each StatsBomb pressure event in the 2025/26 Premier League names the pressing player. We identified the opposing ball carrier from nearby on-ball events and located that player in the same 360 frame as the presser. A pressure could qualify whether or not it belonged to a pressing sequence. Across all 380 fixtures, we retained 20,293 pressures for which we could locate the outfield presser and the carrier, test nearby alternatives, and determine whether the pressing team regained controlled possession in play within five seconds.

We fitted a binary XGBoost classifier to predict this five-second regain outcome, giving each of the 20,293 eligible pressures equal training weight. At each recorded pressure, we calculated the same 26 inputs as for the sequence classifier, using the play preceding that event and constructing each team's shape graph independently (Supplementary Section~\ref{app:model_sensitivities}). We used five-fold cross-validation, keeping whole matches together. Supplementary Section~\ref{app:positioning} gives the carrier-identification procedure, outcome rules, model settings, and validation results.

For the positioning experiment, we selected as \textit{high presses} the 8,607 pressures that met the spatial criteria of Section~\ref{sec:sequence} at the observed positions. The carrier was in their team's own half, the pressure score exceeded 0.65, and at least one visible pressing outfielder was within 6.4 meters of the carrier. At each high press, we moved the recorded presser, who could differ from the defender nearest the carrier. We used a scrambled Sobol sequence to sample 64 candidate positions uniformly by area between 0.5 and 2 meters from the presser's observed position. We retained candidates inside the pitch and the visible area of the frame and at least 0.5 meters from every other visible player. At each valid alternative, we held all other players fixed and rebuilt the structural pressure network. We also held the preceding play fixed, so changes in pressure and space were calculated relative to the same preceding observation at every candidate position (Supplementary Section~\ref{app:positioning_sampling}).

We scored the observed position and its valid alternatives with the same classifier, fitted with that match held out. We defined \textit{model-based position quality} for pressure event $e$ as
\begin{equation}
	Q_e = 100\left[\widehat{p}_e^{\,\mathrm{obs}}-\frac{1}{|\mathcal{A}_e|}\sum_{a\in\mathcal{A}_e}\widehat{p}_e(a)\right],\label{eq:position_response}
\end{equation}
where $\widehat{p}_e^{\,\mathrm{obs}}$ is the predicted regain probability at the observed position, $\mathcal{A}_e$ is the set of valid alternatives, and $\widehat{p}_e(a)$ is the predicted regain probability at alternative $a$. We express $Q_e$ in percentage points. A positive value means the model assigned the observed position a higher regain probability than the mean across its alternatives.

We averaged $Q_e$ within players and clubs, giving each eligible high press equal weight. The player ranking required at least 15 eligible high presses and at least 1,350 Premier League minutes during the season, equivalent to 15 full matches. Club means included all eligible high presses. To obtain 95\% intervals for player means, we resampled each player's contributing fixtures with replacement 2,000 times. We compared club means with final season points using Spearman's rank correlation. For its 95\% interval, we resampled the season's 380 fixtures with replacement 10,000 times, recomputing all club means in each draw and holding final points fixed. Supplementary Section~\ref{app:positioning} gives the matchweek comparison and sampling checks.

We applied the same eligibility criteria, positioning procedure, and five-second regain definition to the 103 matches of the 2026 men's World Cup with 360 data. Because no tournament match belonged to any of the five Premier League folds, we used all five fitted classifiers without refitting and averaged their predicted regain probabilities at each observed or alternative position. We evaluated the predictions at the observed positions against the recorded five-second regain outcomes and used Eq.~\ref{eq:position_response} to calculate position quality for each eligible high press. We ranked players by their mean position quality, requiring at least five eligible high presses and at least 180 tournament minutes (Supplementary Section~\ref{app:positioning_world_cup}).

\section*{Additional Information}

\subsubsection*{Competing interests}\label{sec:competing_interests}

The authors declare no competing interests.

\subsubsection*{Data and code availability}\label{sec:availability}

The data used for the full analysis were supplied under an institutional license from StatsBomb and are subject to its access conditions. The fitted sequence and regain models and the code for applying them are available at \url{https://github.com/Terreturn/Football-pressure-sequence-analysis}, where we show how to use publicly available \href{https://github.com/statsbomb/open-data}{event and 360 data} from the 2022 men's World Cup to construct structural pressure networks, calculate each team's expected press value, and evaluate the positioning of individual pressers. Reproducing the full study requires access to the licensed data. 

\printbibliography[title={References}]

\clearpage

\appendix
\begin{refsection}

\setcounter{figure}{0}
\setcounter{table}{0}
\setcounter{equation}{0}
\renewcommand{\thefigure}{A\arabic{figure}}
\renewcommand{\thetable}{A\arabic{table}}
\renewcommand{\theequation}{A\arabic{equation}}
\renewcommand{\theHfigure}{A\arabic{figure}}
\renewcommand{\theHtable}{A\arabic{table}}
\renewcommand{\theHequation}{A\arabic{equation}}

\phantomsection
\begin{center}
{\Large\bfseries Supplementary Information for}\\[0.35em]
{\large\bfseries \papertitle}\\[0.75em]
{\normalsize Youcheng Song, Maddalena Torricelli, Brennan Klein}\\[0.35em]
{\small \texttt{\url{b.klein@northeastern.edu}}}
\end{center}

\section{Detection, outcomes, and data coverage}\label{app:outcomes}

\subsection{Detecting pressure at an on-ball event}\label{app:detector_details}

The pressure score combines a contribution from each visible opposing outfielder with one from the nearest pitch boundary. An opponent or boundary at distance $d$ contributes
\begin{equation}
	q(d;a,\sigma)=\left[1+\exp\left(\frac{\pi(d-a)}{\sqrt{3}\sigma}\right)\right]^{-1},\label{eq:distance_pressure}
\end{equation}
where $d$ is in meters, $a$ is the distance at which the contribution falls to 0.5, and $\sigma$ controls how quickly it falls with distance. We used $(a,\sigma)=(6.4,2.2)$ meters for outfield opponents and $(3.2,1.1)$ meters for the nearest boundary. Writing $q_j$ for opponent $j$'s contribution and $q_b$ for the boundary contribution, we calculated the score as $1-(1-q_b)\prod_j(1-q_j)$. We required this score to exceed 0.65 and an opposing outfielder to be within 6.4 meters, so a nearby boundary alone could never qualify an event. In the sequence and regain classifiers, direct pressure on the carrier omits the boundary contribution (Table~\ref{tab:features}).

For a dribble, we first looked for the end location of a preceding carry. If that location was unavailable, we used the start location of a following carry, regardless of the dribble outcome. The carry and dribble had to involve the same player, team, and possession and occur in the same period, with no more than 0.01 seconds between the end of one and the start of the other. When a dribble location was recorded, the carry location also had to agree with it within 0.2 provider-grid units. If neither carry supplied a usable location, we used the recorded dribble location when available.

\subsection{Grouping events and assigning an outcome}\label{app:sequence_resolution}

Two consecutive high-pressure events belonged to the same sequence when they occurred in the same period of the match, the same team had the ball for both, and no more than five seconds passed between the end of the earlier event and the start of the later one. The sequence ended if the earlier event or any action between the two went out of play, was ruled offside, or took the ball across halfway. It also ended at an intervening on-ball action by the opponent, a foul, a shot, a substitution or injury stoppage, a restart, a goalkeeper gaining control, or an action by the team in possession that had a frame but did not meet the high-pressure criteria. If an intervening action lacked a frame, we kept the sequence open unless the event record itself showed that the press had ended.

To assign the outcome, we followed the final high-pressure event through the subsequent play. A regain, a pressing-team goalkeeper collection, an opposition clearance that lost possession, a foul won by the pressing team, or a restart awarded to it counted as \textit{success}. A regain required a pass, carry, dribble, or shot by the pressing team, and a recorded ball recovery or interception did not count on its own. A completed escape across halfway, a forward or long lateral escape, a goal scored by the team in possession, or a foul committed by the pressing team counted as \textit{failure}. A retained opposition restart, a return to the opposition goalkeeper, or a release from pressure without an escape counted as \textit{neither}. An out-of-play or offside decision took precedence over an apparent escape across halfway. If the period ended before we could determine the outcome, or the events did not establish one, we excluded the sequence as unresolved.

When the opponent retained possession and the pressure lifted, we classified the release using the direction and length of the ball's movement during the final high-pressure event. If that event was a dribble, or if its pass or carry lacked a start or end location, we left the sequence unresolved. A displacement shorter than 3 meters counted as neither. For displacements of at least 3 meters, movement within 60 degrees of the opponent's attacking direction counted as a forward escape, and a switch of at least 25 meters at an angle greater than 60 and less than 120 degrees from forward counted as a long lateral escape. Both types of escape counted as failure for the pressing team. All other classifiable releases counted as neither.

\subsection{Linking frames to events by location}\label{app:frame_linkage}

Each StatsBomb 360 frame records the identifier of the event it belongs to. In 3,450 of the 11,239 matches with both event and 360 data, none of these identifiers appeared in the event records, while in every other match at least 96.9\% did. The 3,450 matches were all in the fitting pool: 3,428 from La Liga and the Bundesliga between 2020/21 and 2024/25, ten from the Premier League, and twelve from Serie A. In these matches, we went through the frames in order and linked each to the earliest event not yet linked whose location agreed with the frame's actor within 0.051 units on each axis of the $120\times80$ grid. Each search covered up to 60 events with recorded locations, starting up to three such events before the previous link. This linked 94.2\% of the frames in these matches, and we discarded the rest. These matches contributed 34.4\% of the sequences in the fitting sample. Hiding the identifiers in the 2025/26 Bundesliga and La Liga and linking by location instead recovered 98.1\% and 98.3\% of the onsets found with identifiers. The outcome changed for 17 of the 107,232 recovered onsets.

\subsection{From detected sequences to the fitting sample}\label{app:cohort_flow}

Table~\ref{tab:population_flow} follows the data from detected presses to sequences with both a usable network at every event and a known outcome. Across the 24 fitting-pool league-seasons, 9,233 matches had both event and 360 data. We reconstructed 1,245,988 resolved, representable sequences in 9,230 of these matches. The sample used to fit the sequence classifiers contained 1,245,248 sequences from 9,220 matches. It did not include the ten Premier League fixtures whose frames we linked by location (Section~\ref{app:frame_linkage}), which would have added 740 sequences, 0.06\% of the total. Because we required a usable network throughout the sequence, an otherwise usable onset could be lost if a later event could not be represented. Of the 48 detected sequences in the 2025/26 Premier League that could not be represented, two had a usable onset and were excluded only because a later event lacked a usable network. No sequence at the 2026 World Cup was excluded for this reason.

\begingroup
\small
\setlength{\tabcolsep}{2.5pt}

\begin{longtable}{lrrrrr}
\caption{\textbf{Coverage by competition and season.} The match column counts fixtures with both event and 360 data. Among the detected sequences, a sequence is represented if we could construct a network at every high-pressure event. A represented sequence is resolved if we could assign an outcome and unresolved otherwise. Fitting-pool counts include the ten Premier League fixtures that were not in the fitting sample (Section~\ref{app:cohort_flow}).}\label{tab:population_flow}\\
\toprule
Dataset & Matches & Detected & Represented & Resolved & Unresolved \\
\midrule
\endfirsthead
\toprule
Dataset & Matches & Detected & Represented & Resolved & Unresolved \\
\midrule
\endhead
\multicolumn{6}{@{}l}{\textit{Fitting pool}} \\*
Bundesliga 2020/21 & 306 & 40,270 & 40,250 & 39,407 & 843 \\
Bundesliga 2021/22 & 305 & 38,790 & 38,774 & 37,822 & 952 \\
Bundesliga 2022/23 & 306 & 38,722 & 38,710 & 37,913 & 797 \\
Bundesliga 2023/24 & 306 & 41,537 & 41,528 & 40,698 & 830 \\
Bundesliga 2024/25 & 306 & 40,083 & 40,072 & 39,179 & 893 \\
Premier League 2020/21 & 380 & 51,615 & 51,521 & 50,971 & 550 \\
Premier League 2021/22 & 380 & 48,689 & 48,608 & 47,938 & 670 \\
Premier League 2022/23 & 380 & 50,378 & 50,319 & 49,640 & 679 \\
Premier League 2023/24 & 380 & 53,561 & 53,496 & 52,787 & 709 \\
Premier League 2024/25 & 380 & 51,441 & 51,398 & 50,800 & 598 \\
La Liga 2020/21 & 380 & 46,260 & 46,246 & 45,742 & 504 \\
La Liga 2021/22 & 380 & 46,057 & 46,043 & 45,376 & 667 \\
La Liga 2022/23 & 380 & 45,283 & 45,269 & 44,565 & 704 \\
La Liga 2023/24 & 380 & 49,250 & 49,237 & 48,542 & 695 \\
La Liga 2024/25 & 380 & 48,315 & 48,299 & 47,635 & 664 \\
MLS 2021 & 471 & 76,118 & 75,876 & 74,297 & 1,579 \\
MLS 2022 & 489 & 78,809 & 78,563 & 76,989 & 1,574 \\
MLS 2023 & 521 & 85,207 & 85,044 & 83,672 & 1,372 \\
MLS 2024 & 522 & 90,189 & 90,035 & 88,685 & 1,350 \\
Serie A 2020/21 & 380 & 50,631 & 50,600 & 49,937 & 663 \\
Serie A 2021/22 & 380 & 49,615 & 49,598 & 48,754 & 844 \\
Serie A 2022/23 & 381 & 49,670 & 49,644 & 48,936 & 708 \\
Serie A 2023/24 & 380 & 48,684 & 48,670 & 48,074 & 596 \\
Serie A 2024/25 & 380 & 48,278 & 48,268 & 47,629 & 639 \\
\addlinespace[2pt]
\textit{Fitting-pool total} & 9,233 & 1,267,452 & 1,266,068 & 1,245,988 & 20,080 \\
\midrule
\multicolumn{6}{@{}l}{\textit{Outside datasets}} \\*
Premier League 2025/26 & 380 & 63,411 & 63,363 & 62,375 & 988 \\
La Liga 2025/26 & 376 & 61,804 & 61,781 & 60,928 & 853 \\
Bundesliga 2025/26 & 300 & 49,373 & 49,358 & 48,229 & 1,129 \\
Serie A 2025/26 & 379 & 60,253 & 60,222 & 59,200 & 1,022 \\
Ligue 1 2025/26 & 302 & 49,162 & 49,152 & 48,087 & 1,065 \\
Euro 2020 & 51 & 6,361 & 6,359 & 6,251 & 108 \\
Euro 2024 & 51 & 6,499 & 6,496 & 6,391 & 105 \\
World Cup 2022 & 64 & 8,060 & 8,053 & 7,922 & 131 \\
World Cup 2026 & 103 & 14,929 & 14,923 & 14,647 & 276 \\
\addlinespace[2pt]
\textit{Outside total} & 2,006 & 319,852 & 319,707 & 314,030 & 5,677 \\
\bottomrule
\end{longtable}

\endgroup

In 2025/26, each of the four European leagues in the fitting pool had more detected pressing sequences per match than in any of its earlier seasons (Table~\ref{tab:population_flow}). In the Premier League, for example, the number rose from 135.4 per match in 2024/25 to 166.9 in 2025/26. We also detected more pressing sequences per match at the 2026 World Cup than at the 2022 edition.

\section{Network feature definitions}\label{app:network}

We built each event's structural pressure network from the same 360 frame that qualified it as a high-pressure event. The network has four layers: pressure links from the pressing outfielders to the carrier and possible receivers, passing links from the carrier to those receivers, a Voronoi cell for each visible player, and a shape graph for each team. The 26 inputs in Table~\ref{tab:features} summarize these layers together with the carrier's position and recent play.

We write $c$ for the carrier, $r$ for a visible teammate who could receive a pass, and $j$ for a visible pressing outfielder. The team in possession attacks toward increasing $x$, so receiver $r$ is ahead of the ball if $x_r>x_c$. The event record and the 360 frame each give a location for the carrier. For sequence events, we built the network from the event location when it was available and took the carrier-position input from the frame's actor.

For each pressing outfielder $j$ and target $t$ (the carrier or a possible receiver), we calculated the pressure contribution $a_{jt}=q(d_{jt};6.4,2.2)$ from their separation $d_{jt}$ in meters (Eq.~\ref{eq:distance_pressure}). We call a link active when $a_{jt}>0.05$. Total pressure on the carrier included every visible pressing outfielder. To measure how pressure was spread across targets, we summed the active contributions at each target, divided each target's sum by total active pressure, and divided the entropy of the resulting distribution by $\log(\max(2,K))$, where $K$ is the number of targets (the carrier plus visible receivers). For presser focus across targets, we applied the same calculation to each presser's active links and averaged over pressers with at least one active link. Larger values of either entropy mean that pressure was spread more evenly. Both were zero when no link was active.

The weight of a passing link is close to 1 for a short pass along a clear line, and the weight decreases as the pass gets longer or as a pressing outfielder stands closer to its line. We calculated it as $w_{cr}=\exp(-d_{cr}/18)(1-b_{cr})$, where $d_{cr}$ is the distance from the carrier to the receiver in meters and $b_{cr}$ measures obstruction. Under this construction, the distance factor is 0.5 for a receiver 12.5 meters away, and therefore, an unobstructed receiver counts as an open passing option ($w_{cr}>0.5$) only when closer than 12.5 meters. To measure obstruction, we considered each pressing outfielder located between the carrier and the receiver along the line of the pass, meaning that their perpendicular projection fell on the segment joining the two players. For an outfielder at perpendicular distance $h$ meters from the line, we calculated $q(h;1.2,0.6)$ using Eq.~\ref{eq:distance_pressure}. We set $b_{cr}$ to the largest of these values, which comes from the outfielder closest to the line, or to zero if that value was 0.05 or less. This value is 0.97 for an outfielder standing on the line, which cuts the passing weight by 97\%. It falls to 0.5 at 1.2 meters from the line and below 0.05 beyond about 2.2 meters. The goalkeeper of the team in possession could be a receiver, and the pressing goalkeeper counted toward neither pressure nor obstruction.

We computed each player's Voronoi cell with both goalkeepers included when visible and clipped it to the part of the pitch in the broadcast camera's view. We write $A_i$ for player $i$'s clipped cell area and $A$ for the visible pitch area. Team shapes are the sparse subgraphs of each team's Delaunay triangulation described by Brandes et al.~\cite{Brandes2025_shape}. We left goalkeepers out of the shape graphs unless the goalkeeper was the carrier.

Several inputs compare the current frame with preceding play. At sequence onset, we searched up to ten preceding event records in the same period, stopping at any stoppage, and selected the most recent one with a usable ball location distinct from the current one. Later in a sequence, incoming-ball distance and direction used the most recent distinct ball location within the sequence and fell back on the selected pre-sequence event when necessary. For the four change inputs, we subtracted the value at the immediately preceding sequence event, even when the ball had stayed at the same location. At onset, we used the value at the selected pre-sequence event when its frame supplied the required measurement. Changes kept each feature's original units and were missing when the earlier value was unavailable.

When no receiver was ahead of the carrier, the best forward passing option and best forward-receiver relief were zero, as was forward passing space when the spatial geometry was available. An area share was missing when we lacked the visible-area polygon or a required Voronoi cell. Local presser spread required at least five visible pressing outfielders, nearest-presser support required at least one shape edge connected to that presser, and the forward share of carrier space required a carrier cell with positive area.

\begingroup
\footnotesize
\setlength{\tabcolsep}{4pt}
\setlength{\LTcapwidth}{\linewidth}

\begin{longtable}{@{}>{\raggedright\arraybackslash}p{0.27\linewidth}>{\raggedright\arraybackslash}p{0.56\linewidth}>{\raggedright\arraybackslash}p{0.12\linewidth}@{}}
\caption{\textbf{Inputs to the pooled sequence classifier.} The nine inputs in the first group describe the play at the ball and also enter the nine-input classifier. The remaining 17 describe the surrounding players.}\label{tab:features}\\
\toprule
Feature & Definition & Units \\
\midrule
\endfirsthead
\toprule
Feature & Definition & Units \\
\midrule
\endhead
\multicolumn{3}{@{}l}{\textbf{Nine-input and full classifiers (9 inputs)}} \\*
Incoming-ball distance & Distance from the selected preceding ball location to the current ball location. & Meters \\
Incoming-ball direction & Cosine of the angle between that displacement and the attacking direction. & Unitless \\
Carrier position & Longitudinal coordinate of the frame's actor divided by pitch length, clipped to $[0,1]$. & Fraction \\
Change in carrier position & Current minus preceding normalized longitudinal position. & Fraction \\
Distance to touchline & Carrier distance to the nearer touchline. & Meters \\
Pressure on carrier & $1-\prod_j(1-a_{jc})$ over visible pressing outfielders. & Unitless \\
Change in carrier pressure & Current minus preceding pressure on the carrier. & Unitless \\
Effective pressers & $(\sum_j a_{jc})^2/\sum_j a_{jc}^2$ over active links to the carrier, or zero with none. & Effective count \\
Recent opponent possession & One if the pressing team performed the selected pre-sequence event, zero if the team in possession performed it, and missing if no event was selected. & Binary \\
\midrule
\multicolumn{3}{@{}l}{\textbf{Full classifier only (17 collective inputs)}} \\*
Open passing options & Number of receivers with passing weight $w_{cr}>0.5$. & Count \\
Best forward passing option & Largest $w_{cr}$ over receivers ahead of the carrier. & Unitless \\
Pressure spread across targets & Normalized entropy of total active pressure across the carrier and all receivers. & Unitless \\
Mean receiver pressure & Mean over receivers of the sum of pressure contributions $a_{jr}\geq0.05$. & Unitless \\
Presser focus across targets & Mean normalized entropy of each active presser's pressure across targets. Higher values mean individual pressers spread their pressure more evenly. & Unitless \\
Carrier space & Carrier's Voronoi area as a share of the visible pitch area, $A_c/A$. & Fraction \\
Forward-receiver space & Sum of $A_r/A$ over receivers ahead of the carrier. & Fraction \\
Change in carrier space & Current minus preceding carrier area share. & Fraction \\
Change in forward-receiver space & Current minus preceding forward-receiver area share. & Fraction \\
Uncrossed shape-graph outlets & Number of outfield receivers whose segment from the carrier crosses no pressing-team shape edge. Touching an endpoint alone does not count as crossing. & Count \\
Opponent shape links cut & Fraction of the possession team's shape edges with a pressing outfielder strictly within 2 meters whose perpendicular projection falls on the edge. & Fraction \\
Longest pressing-team shape link & Maximum edge length in the pressing team's shape graph. & Meters \\
Best forward-receiver relief & Largest $w_{cr}\prod_j(1-a_{jr})$ over forward receivers, combining an open lane with lower pressure at the receiver. & Unitless \\
Forward passing space & Sum of $w_{cr} A_r/A$ over forward receivers. & Fraction \\
Local presser spread & Mean distance of the five pressing outfielders nearest the carrier from their centroid. & Meters \\
Nearest-presser support & Median length of the pressing-team shape edges incident to the presser nearest the carrier. & Meters \\
Forward share of carrier space & Fraction of the carrier's clipped Voronoi cell lying ahead of its longitudinal position. & Fraction \\
\bottomrule
\end{longtable}

\endgroup

\subsection{Building the team shape graphs}\label{app:model_sensitivities}

For the sequence classifiers, we recorded the uncrossed shape-graph outlets, opponent shape links cut, longest pressing-team shape link, and nearest-presser support as missing if either team's shape graph could not be constructed. For the regain model, we constructed the two teams' graphs separately, so a failure for one team left the inputs based on the other team's graph available.

\section{Model specification and prediction comparisons}\label{app:model}

We fitted a three-outcome XGBoost classifier~\cite{Chen2016_xgboost} to the 26 inputs in Table~\ref{tab:features}. We used 2,850 boosting rounds with XGBoost's histogram method, maximum depth 7, learning rate 0.05, minimum child weight 3.7355, minimum loss reduction for a split 1.44164, and $L_2$ penalty 3.80847. Row and column sampling fractions were 0.882904 and 0.871906, respectively, and the random seed was 202405. We used these settings for the full fitting pool, for each cross-validation fold, and for the nine-input classifier. During model development, we chose the maximum depth and minimum child weight by comparing depths of 7, 9, and 11 and two minimum child weights, with early stopping, on a fixed sample of 1,383 fitting-pool matches (15\%) stratified by league-season. This sample included matches from all five cross-validation folds and none from the outside datasets.

Each event in a resolved sequence was labeled with the sequence's outcome and weighted by $1/n$, where $n$ is the number of events in that sequence. We used five folds of whole matches without shuffling (\texttt{GroupKFold}). During model development, we examined performance in the outside datasets when comparing candidate features. For the outside predictions in Fig.~\ref{fig:prediction}, we fitted the sequence classifier and estimated its calibration temperature using only the fitting pool (Section~\ref{sec:methods_model}). Table~\ref{tab:outside_auc} gives the success AUC and interval for each dataset in Fig.~\ref{fig:prediction}a.

Presses from the same match share players and opponents. For success AUC, calibration comparisons, and the main log-loss comparison, we therefore sampled whole matches with replacement, taking every evaluated onset in each selected match. We drew 200 samples within competition-season and, in the fitting pool, within classifier fold and calibration half (Section~\ref{app:calibration}). For comparisons between classifiers, we used the same sampled matches for both. We held the fitted trees fixed and refitted the calibration parameters in each draw using the sampled fitting-pool predictions, including when evaluating an outside dataset. We took the 2.5th and 97.5th percentiles as the limits of a 95\% interval. These intervals describe sampling variation across fixtures conditional on the fitted trees and the competitions studied.

\begingroup
\footnotesize
\setlength{\tabcolsep}{4pt}
\setlength{\LTcapwidth}{\linewidth}

\begin{longtable}{@{}lrrrl@{}}
\caption{\textbf{Discrimination at sequence onset.} AUC measures how well predicted success probabilities rank sequences that succeeded above those ending in failure or neither. Intervals are 95\% match-bootstrap intervals. Each outside dataset used the same fitted classifier and temperature.}\label{tab:outside_auc}\\
\toprule
Dataset & Matches & Onsets & AUC & 95\% interval \\
\midrule
\endfirsthead
\toprule
Dataset & Matches & Onsets & AUC & 95\% interval \\
\midrule
\endhead
Cross-validation & 9,220 & 1,245,248 & $0.7092$ & $[0.7079,\,0.7104]$ \\
Premier League 2025/26 & 380 & 62,375 & $0.7278$ & $[0.7232,\,0.7325]$ \\
La Liga 2025/26 & 376 & 60,928 & $0.7200$ & $[0.7154,\,0.7252]$ \\
Bundesliga 2025/26 & 300 & 48,229 & $0.7333$ & $[0.7283,\,0.7393]$ \\
Serie A 2025/26 & 379 & 59,200 & $0.7316$ & $[0.7262,\,0.7361]$ \\
Ligue 1 2025/26 & 302 & 48,087 & $0.7238$ & $[0.7182,\,0.7299]$ \\
Euro 2020 & 51 & 6,251 & $0.7498$ & $[0.7283,\,0.7713]$ \\
Euro 2024 & 51 & 6,391 & $0.7412$ & $[0.7249,\,0.7575]$ \\
World Cup 2022 & 64 & 7,922 & $0.7145$ & $[0.6992,\,0.7256]$ \\
World Cup 2026 & 103 & 14,647 & $0.7282$ & $[0.7195,\,0.7383]$ \\
\bottomrule
\end{longtable}

\endgroup

\subsection{Additional information from network features}\label{app:collective}

We fitted the nine- and 26-input classifiers as described in Section~\ref{sec:methods_comparison} and compared them at the same onsets. For $N$ evaluated onsets, let $y_s$ denote the observed outcome of sequence $s$ and $p_{s,y_s}$ the calibrated probability assigned to that outcome at onset. The multiclass log loss is
\begin{equation}
	\mathcal{L}=-\frac{1}{N}\sum_{s=1}^{N}\log p_{s,y_s},\label{eq:onset_log_loss}
\end{equation}
where the logarithm is natural and the loss is in nats per onset. We report the absolute reduction $\mathcal{L}_{9}-\mathcal{L}_{26}$ and the relative reduction $(\mathcal{L}_{9}-\mathcal{L}_{26})/\mathcal{L}_{9}$, weighting each onset equally in pooled estimates. Table~\ref{tab:collective_comparison} reports both losses, their difference, and a paired 95\% match-bootstrap interval for the difference (Section~\ref{app:model}).

\begingroup
\footnotesize
\setlength{\tabcolsep}{4pt}
\setlength{\LTcapwidth}{\linewidth}

\begin{longtable}{@{}lrrrl@{}}
\caption{\textbf{Added information from 17 collective features.} Onset log loss for the nine-input and 26-input classifiers, in nats per onset. Positive reductions favor the full classifier. Intervals are paired 95\% match-bootstrap intervals.}\label{tab:collective_comparison}\\
\toprule
Dataset & 9 inputs & 26 inputs & Reduction & 95\% interval \\
\midrule
\endfirsthead
\toprule
Dataset & 9 inputs & 26 inputs & Reduction & 95\% interval \\
\midrule
\endhead
Cross-validation & $0.9700$ & $0.9434$ & $0.0265$ & $[0.0262,\,0.0269]$ \\
Premier League 2025/26 & $1.0051$ & $0.9763$ & $0.0287$ & $[0.0270,\,0.0307]$ \\
La Liga 2025/26 & $1.0019$ & $0.9753$ & $0.0266$ & $[0.0248,\,0.0285]$ \\
Bundesliga 2025/26 & $1.0089$ & $0.9781$ & $0.0308$ & $[0.0283,\,0.0330]$ \\
Serie A 2025/26 & $0.9961$ & $0.9662$ & $0.0299$ & $[0.0278,\,0.0321]$ \\
Ligue 1 2025/26 & $0.9909$ & $0.9645$ & $0.0265$ & $[0.0246,\,0.0285]$ \\
Euro 2020 & $0.9325$ & $0.9016$ & $0.0309$ & $[0.0252,\,0.0367]$ \\
Euro 2024 & $0.9331$ & $0.8992$ & $0.0339$ & $[0.0286,\,0.0391]$ \\
World Cup 2022 & $0.9455$ & $0.9114$ & $0.0341$ & $[0.0302,\,0.0388]$ \\
World Cup 2026 & $0.9953$ & $0.9649$ & $0.0305$ & $[0.0266,\,0.0338]$ \\
Outside pooled & $0.9963$ & $0.9674$ & $0.0289$ & $[0.0280,\,0.0296]$ \\
\bottomrule
\end{longtable}

\endgroup

Some of the information in the collective features could come from how many players the frame showed or how many defenders were near the carrier. To test this, we added eleven local-defender and frame-count inputs to the nine-input classifier: the distance and the sine and cosine of the direction of each of the two nearest pressing outfielders, the numbers of pressing outfielders within 5 and 10 meters of the carrier, and three counts of visible players (all players, pressing outfielders, and the carrier plus outfield receivers). We compared this 20-input classifier with a 37-input classifier that also included the 17 collective features. We also grouped matches by how often a recorded possession began with a block or clearance, using this frequency as a proxy for differences in possession recording. We fitted and calibrated both classifiers separately for matches in which at least 1.5\% of possessions in the first two periods began this way and for those below that threshold. Adding the collective features still reduced onset log loss, by 0.0183 nats per onset in cross-validation and 0.0190 nats per onset across the pooled outside onsets.

\subsection{TreeSHAP contributions}\label{app:shap}

We computed exact TreeSHAP contributions~\cite{Lundberg2020_treeshap} at all 314,030 outside onsets for the 26-input classifier fitted to the whole fitting pool. We used XGBoost's tree-path-dependent calculation, which weights alternative paths by the training cover stored at tree nodes. At each onset, the baseline and the 26 signed contributions sum to the classifier's raw score for each outcome. We divided both by the fitted temperature, which puts them on the scale of the calibrated scores. We then ranked the inputs by their mean absolute contribution across the three calibrated outcome scores and all outside onsets, weighting each outcome and each onset equally (Table~\ref{tab:shap}). For the 95\% intervals, we resampled whole matches with replacement within each outside dataset 500 times, holding the classifier and its temperature fixed.

\begingroup
\footnotesize
\setlength{\tabcolsep}{4pt}
\setlength{\LTcapwidth}{\linewidth}

\begin{longtable}{@{}rllrl@{}}
\caption{\textbf{Feature contributions at sequence onset.} Mean absolute TreeSHAP contributions across the three calibrated outcome scores and all outside onsets, in class-score units before conversion to probabilities. Group separates the nine inputs describing play at the ball, which both classifiers used, from the 17 collective features (Table~\ref{tab:features}). We obtained 95\% intervals for the pooled contributions by resampling matches within each outside dataset.}\label{tab:shap}\\
\toprule
Rank & Feature & Group & Mean $|\mathrm{SHAP}|$ & 95\% interval \\
\midrule
\endfirsthead
\toprule
Rank & Feature & Group & Mean $|\mathrm{SHAP}|$ & 95\% interval \\
\midrule
\endhead
1 & Carrier position & Ball & $0.2194$ & $[0.2189,\,0.2199]$ \\
2 & Pressure spread across targets & Collective & $0.0821$ & $[0.0818,\,0.0824]$ \\
3 & Incoming-ball direction & Ball & $0.0699$ & $[0.0698,\,0.0701]$ \\
4 & Distance to touchline & Ball & $0.0669$ & $[0.0667,\,0.0671]$ \\
5 & Uncrossed shape-graph outlets & Collective & $0.0648$ & $[0.0645,\,0.0651]$ \\
6 & Best forward-receiver relief & Collective & $0.0615$ & $[0.0614,\,0.0617]$ \\
7 & Change in carrier position & Ball & $0.0594$ & $[0.0592,\,0.0595]$ \\
8 & Pressure on carrier & Ball & $0.0592$ & $[0.0590,\,0.0593]$ \\
9 & Best forward passing option & Collective & $0.0442$ & $[0.0441,\,0.0442]$ \\
10 & Recent opponent possession & Ball & $0.0425$ & $[0.0423,\,0.0426]$ \\
11 & Change in carrier pressure & Ball & $0.0402$ & $[0.0401,\,0.0403]$ \\
12 & Forward share of carrier space & Collective & $0.0379$ & $[0.0378,\,0.0380]$ \\
13 & Presser focus across targets & Collective & $0.0336$ & $[0.0335,\,0.0337]$ \\
14 & Incoming-ball distance & Ball & $0.0320$ & $[0.0318,\,0.0321]$ \\
15 & Mean receiver pressure & Collective & $0.0297$ & $[0.0295,\,0.0298]$ \\
16 & Local presser spread & Collective & $0.0292$ & $[0.0291,\,0.0293]$ \\
17 & Open passing options & Collective & $0.0290$ & $[0.0289,\,0.0291]$ \\
18 & Forward passing space & Collective & $0.0287$ & $[0.0286,\,0.0288]$ \\
19 & Change in forward-receiver space & Collective & $0.0264$ & $[0.0263,\,0.0265]$ \\
20 & Forward-receiver space & Collective & $0.0239$ & $[0.0238,\,0.0240]$ \\
21 & Change in carrier space & Collective & $0.0192$ & $[0.0191,\,0.0192]$ \\
22 & Carrier space & Collective & $0.0186$ & $[0.0185,\,0.0186]$ \\
23 & Effective pressers & Ball & $0.0180$ & $[0.0180,\,0.0181]$ \\
24 & Nearest-presser support & Collective & $0.0179$ & $[0.0179,\,0.0180]$ \\
25 & Opponent shape links cut & Collective & $0.0124$ & $[0.0124,\,0.0124]$ \\
26 & Longest pressing-team shape link & Collective & $0.0109$ & $[0.0109,\,0.0110]$ \\
\bottomrule
\end{longtable}

\endgroup

\subsection{Probability calibration}\label{app:calibration}

To test whether adjusting the three outcome scores improved agreement with observed outcomes~\cite{Guo2017_calibration}, we compared the original probabilities with three adjustment rules: a common temperature, which changes how strongly the predicted probabilities favor the highest-scoring outcome; separate offsets for success and failure, which shift their average predictions relative to neither; and both together. We fitted each rule by minimizing multiclass log loss with equal weight per onset, using either all fitting-pool seasons or, to check whether recent matches gave a better calibration, only the 243,150 onsets in 1,785 matches played from May 25, 2024 through May 25, 2025. To evaluate each rule on unseen matches, we split the matches within each held-out classifier fold and competition-season into two halves using a reproducible hash of their identifiers, estimated the rule on one half, scored the other, and then exchanged their roles.

We preferred a simpler rule when the paired 95\% interval for its difference from the lowest-loss rule included zero. A common temperature fitted to all seasons gave a held-out loss 0.000012 nats per onset higher than temperature and offsets together (paired 95\% interval $[-0.000047,\,0.000032]$), so we kept the common temperature. In the outside datasets it also gave the lowest loss of the seven options (the original probabilities and each rule fitted to all seasons or to recent matches), 0.9674 nats per onset against 0.9682 to 0.9705 for the others. To score the outside datasets, we estimated this temperature once from held-out predictions at all 1,245,248 fitting-pool onsets.

For onset $s$ with uncalibrated probabilities $p_{sk}$ for outcome $k$, temperature scaling gives
\begin{equation}
	\widetilde{p}_{sk}=\frac{p_{sk}^{1/T}}{\sum_{\ell\in\{-,0,+\}}p_{s\ell}^{1/T}},\label{eq:temperature_calibration}
\end{equation}
where $T$ is shared across the three outcomes. The value used outside the fitting pool was $T\approx1.099$ ($1/T=0.909793$), which draws extreme probabilities slightly toward the center. We estimated the success calibration slope by logistic regression of the observed success indicator on the predicted log odds of success. In cross-validation, this slope moved from 0.912 before calibration to 1.005 after calibration, close to the value of one expected when the predicted spread agrees with observations. Across the five outside league-seasons, mean predicted success was 19.42\% versus 24.20\% observed, and mean predicted failure was 42.60\% versus 38.72\% observed, leaving the predicted success-minus-failure balance 8.66 percentage points below the observed balance (Table~\ref{tab:calibration}). The bias depended on the season: success was under-predicted in all six 2025/26 datasets, including the 2026 World Cup, and over-predicted in the three earlier tournaments.

\begingroup
\footnotesize
\setlength{\tabcolsep}{3.5pt}
\setlength{\LTcapwidth}{\linewidth}

\begin{longtable}{@{}lrrrrrlr@{}}
\caption{\textbf{Onset calibration of the sequence classifier.} Mean predicted and observed percentages of sequences ending in success and failure; the rest ended in neither. Balance bias is the predicted balance (success minus failure) minus the observed balance, in percentage points, with its 95\% match-bootstrap interval. Slope is the calibration slope for success.}\label{tab:calibration}\\
\toprule
 & \multicolumn{2}{c}{Success (\%)} & \multicolumn{2}{c}{Failure (\%)} & \multicolumn{2}{c}{Balance bias (pp)} & \\
\cmidrule(lr){2-3}\cmidrule(lr){4-5}\cmidrule(lr){6-7}
Sample & Pred. & Obs. & Pred. & Obs. & Bias & 95\% interval & Slope \\
\midrule
\endfirsthead
\toprule
 & \multicolumn{2}{c}{Success (\%)} & \multicolumn{2}{c}{Failure (\%)} & \multicolumn{2}{c}{Balance bias (pp)} & \\
\cmidrule(lr){2-3}\cmidrule(lr){4-5}\cmidrule(lr){6-7}
Sample & Pred. & Obs. & Pred. & Obs. & Bias & 95\% interval & Slope \\
\midrule
\endhead
Cross-validation & 18.37 & 18.13 & 42.75 & 42.99 & 0.47 & $[0.37,\,0.60]$ & 1.005 \\
Premier League 2025/26 & 19.36 & 24.95 & 42.55 & 37.90 & $-10.24$ & $[-10.83,\,-9.72]$ & 1.082 \\
La Liga 2025/26 & 19.48 & 23.89 & 42.99 & 39.79 & $-7.61$ & $[-8.40,\,-6.97]$ & 1.068 \\
Bundesliga 2025/26 & 20.14 & 25.88 & 42.65 & 37.71 & $-10.68$ & $[-11.33,\,-9.85]$ & 1.096 \\
Serie A 2025/26 & 18.89 & 23.83 & 42.67 & 39.18 & $-8.42$ & $[-9.00,\,-7.72]$ & 1.091 \\
Ligue 1 2025/26 & 19.34 & 22.37 & 42.06 & 38.86 & $-6.23$ & $[-7.00,\,-5.69]$ & 1.057 \\
Euro 2020 & 17.07 & 13.66 & 42.97 & 43.74 & 4.18 & $[2.38,\,6.26]$ & 1.176 \\
Euro 2024 & 17.70 & 14.40 & 41.71 & 40.35 & 1.95 & $[0.17,\,3.52]$ & 1.091 \\
World Cup 2022 & 16.95 & 14.33 & 42.01 & 42.88 & 3.49 & $[1.73,\,5.47]$ & 1.011 \\
World Cup 2026 & 18.65 & 22.39 & 42.55 & 38.89 & $-7.40$ & $[-9.03,\,-5.68]$ & 1.062 \\
\midrule
Five outside leagues & 19.42 & 24.20 & 42.60 & 38.72 & $-8.66$ & $[-8.95,\,-8.31]$ & 1.080 \\
Outside pooled & 19.24 & 23.45 & 42.57 & 38.96 & $-7.83$ & $[-8.12,\,-7.50]$ & 1.083 \\
\bottomrule
\end{longtable}

\endgroup

\subsection{Player pressing efficiency at the 2026 men's World Cup}\label{app:world_cup_efficiency}

For the pressing-efficiency ranking in Table~\ref{tab:world_cup_players}, we used a separate 15-input classifier fitted to events from the 2026 men's World Cup using XGBoost 3.0.0. Each event received its own outcome label and equal training weight. We applied the fitted classifier to 30,381 events from the 103 tournament fixtures with 360 data and used its predicted probabilities without further calibration. At the first and last scored events of each sequence, we subtracted the predicted probability of failure from the predicted probability of success. The increase or decrease in this difference over the sequence, expressed in percentage points, gave the sequence its score.

We considered sequences with at least two scored events, all in the same period and possession. We linked a player to a sequence if they applied an eligible high press for the pressing team during that possession and period, with the pressure recorded between the first and last scored events in event order, inclusive. Each sequence counted once per player. Teammates linked to the same sequence received its full score. We averaged these scores for each player and ranked the 17 players with at least five linked sequences and at least 180 tournament minutes.

\section{Team measures and performance associations}\label{app:teams}

The team analysis covered 599 team-seasons across 29 league-seasons. We used 1,524,067 eligible onsets to estimate expected press value and 10,985 fixtures with performance data to calculate season results. We assigned each sequence to the pressing team and averaged the calibrated onset probability of success minus that of failure within each team-season (Eq.~\ref{eq:team_expected_press_value}). In Fig.~\ref{fig:team_value}a, zero is the mean over eligible presses in the same league-season. Subtracting this common value preserves the ordering of teams.

We summed shot expected goals for and against each team, including extra time and excluding shoot-outs, and divided the difference by the number of fixtures with performance data. We also calculated points per match from recorded scores, assigning three points for a win and one for a draw. We included MLS playoff fixtures and excluded one 2021/22 Bundesliga fixture without a recorded score. Table~\ref{tab:team_xgd} gives the within-season Spearman correlations with both measures and the approximate 95\% intervals. Expected press value correlated positively with expected-goal difference per match in all 29 league-seasons (median $\rho=0.52$, range 0.16 to 0.81), with 95\% intervals above zero in 17 of them. Its median correlation with points per match was 0.43, and that correlation was positive in every league-season, although only 0.002 in MLS 2021 (shown as 0.00 in Table~\ref{tab:team_xgd}).

For the correlations across teams in Fig.~\ref{fig:team_value} and Table~\ref{tab:team_xgd}, each observation was a team's season average. We calculated approximate 95\% intervals with the Bonett-Wright approximation~\cite{Bonett2000_correlations}. For $n$ teams and estimated Spearman correlation $\rho$, the standard error on the Fisher scale is
\begin{equation}
	\operatorname{SE}\!\left[\operatorname{atanh}(\rho)\right]=\sqrt{\frac{1+\rho^2/2}{n-3}}.\label{eq:spearman_uncertainty}
\end{equation}
We transformed $\operatorname{atanh}(\rho)\pm1.96\operatorname{SE}$ back to the correlation scale to obtain the interval limits. This approximation treats clubs as independent within a league-season.

To examine where teams began their presses, we averaged the carrier's longitudinal position over the onsets of each team's presses. We ranked teams by this mean position and by expected-goal difference per match within the same league-season. Because the carrier's coordinate increases with distance from the carrier's own goal line, a smaller value places the press closer to the carrier's own goal, which the pressing team is attacking. Spearman's correlation was negative in all 29 league-seasons (median $\rho=-0.67$, range $-0.89$ to $-0.29$). Teams with higher expected-goal difference thus tended to begin their presses closer to that goal. Estimating how much starting position explains the association between expected press value and expected-goal difference would require an adjusted comparison.

Temperature scaling had little effect on team rankings. The median within-season Spearman correlation between rankings with and without it was 0.997. Of the 599 team-seasons, 486 kept their rank, 585 stayed within one place, and none moved more than three places.

\begingroup
\footnotesize
\setlength{\tabcolsep}{4pt}
\setlength{\LTcapwidth}{\linewidth}

\begin{longtable}{@{}lrrlrl@{}}
\caption{\textbf{Expected press value and season performance.} Within-season Spearman correlations with expected-goal difference (xGD) and points per match. Intervals use the Bonett-Wright approximation. Daggers mark the five outside league-seasons.}\label{tab:team_xgd}\label{tab:team_points}\\
\toprule
League-season & Teams & xGD $\rho$ & 95\% interval & Points $\rho$ & 95\% interval \\
\midrule
\endfirsthead
\toprule
League-season & Teams & xGD $\rho$ & 95\% interval & Points $\rho$ & 95\% interval \\
\midrule
\endhead
Bundesliga 2020/21 & 18 & $0.74$ & $[0.36,\,0.91]$ & $0.75$ & $[0.39,\,0.91]$ \\
Bundesliga 2021/22 & 18 & $0.39$ & $[-0.11,\,0.73]$ & $0.43$ & $[-0.07,\,0.76]$ \\
Bundesliga 2022/23 & 18 & $0.28$ & $[-0.22,\,0.67]$ & $0.05$ & $[-0.43,\,0.50]$ \\
Bundesliga 2023/24 & 18 & $0.54$ & $[0.06,\,0.81]$ & $0.50$ & $[0.02,\,0.80]$ \\
Bundesliga 2024/25 & 18 & $0.64$ & $[0.21,\,0.87]$ & $0.58$ & $[0.11,\,0.84]$ \\
Bundesliga 2025/26$^\dagger$ & 18 & $0.76$ & $[0.39,\,0.92]$ & $0.81$ & $[0.49,\,0.94]$ \\
Premier League 2020/21 & 20 & $0.55$ & $[0.11,\,0.81]$ & $0.40$ & $[-0.07,\,0.72]$ \\
Premier League 2021/22 & 20 & $0.67$ & $[0.28,\,0.87]$ & $0.50$ & $[0.05,\,0.78]$ \\
Premier League 2022/23 & 20 & $0.60$ & $[0.17,\,0.84]$ & $0.59$ & $[0.16,\,0.83]$ \\
Premier League 2023/24 & 20 & $0.52$ & $[0.07,\,0.79]$ & $0.45$ & $[-0.02,\,0.75]$ \\
Premier League 2024/25 & 20 & $0.58$ & $[0.14,\,0.82]$ & $0.47$ & $[0.01,\,0.77]$ \\
Premier League 2025/26$^\dagger$ & 20 & $0.67$ & $[0.28,\,0.87]$ & $0.59$ & $[0.16,\,0.83]$ \\
La Liga 2020/21 & 20 & $0.58$ & $[0.15,\,0.83]$ & $0.24$ & $[-0.23,\,0.62]$ \\
La Liga 2021/22 & 20 & $0.40$ & $[-0.08,\,0.72]$ & $0.20$ & $[-0.27,\,0.59]$ \\
La Liga 2022/23 & 20 & $0.47$ & $[0.01,\,0.76]$ & $0.30$ & $[-0.17,\,0.66]$ \\
La Liga 2023/24 & 20 & $0.25$ & $[-0.22,\,0.63]$ & $0.45$ & $[-0.01,\,0.76]$ \\
La Liga 2024/25 & 20 & $0.55$ & $[0.10,\,0.81]$ & $0.43$ & $[-0.04,\,0.74]$ \\
La Liga 2025/26$^\dagger$ & 20 & $0.33$ & $[-0.15,\,0.68]$ & $0.14$ & $[-0.33,\,0.55]$ \\
Ligue 1 2025/26$^\dagger$ & 18 & $0.81$ & $[0.50,\,0.94]$ & $0.80$ & $[0.47,\,0.93]$ \\
MLS 2021 & 27 & $0.48$ & $[0.10,\,0.74]$ & $0.00$ & $[-0.38,\,0.38]$ \\
MLS 2022 & 28 & $0.26$ & $[-0.13,\,0.58]$ & $0.17$ & $[-0.22,\,0.51]$ \\
MLS 2023 & 29 & $0.20$ & $[-0.18,\,0.53]$ & $0.14$ & $[-0.24,\,0.49]$ \\
MLS 2024 & 29 & $0.16$ & $[-0.22,\,0.50]$ & $0.11$ & $[-0.27,\,0.46]$ \\
Serie A 2020/21 & 20 & $0.67$ & $[0.27,\,0.87]$ & $0.74$ & $[0.40,\,0.90]$ \\
Serie A 2021/22 & 20 & $0.40$ & $[-0.08,\,0.72]$ & $0.37$ & $[-0.11,\,0.70]$ \\
Serie A 2022/23 & 20 & $0.45$ & $[-0.02,\,0.75]$ & $0.48$ & $[0.02,\,0.77]$ \\
Serie A 2023/24 & 20 & $0.26$ & $[-0.22,\,0.63]$ & $0.20$ & $[-0.27,\,0.60]$ \\
Serie A 2024/25 & 20 & $0.36$ & $[-0.11,\,0.70]$ & $0.32$ & $[-0.15,\,0.68]$ \\
Serie A 2025/26$^\dagger$ & 20 & $0.56$ & $[0.12,\,0.82]$ & $0.69$ & $[0.31,\,0.88]$ \\
\bottomrule
\end{longtable}

\endgroup

\section{Counterfactual positioning experiments}\label{app:positioning}

\subsection{Pressure-event eligibility and carrier identification}\label{app:positioning_cohort}

At a recorded pressure event, the actor named in the event record is the presser. To score that defender's position, we also needed to identify the opposing ball carrier and locate that player among the opponents in the pressure frame. We required one outfield actor at the recorded pressure location, at least four player nodes, and a valid polygon marking the visible area. We considered each recorded pressure separately, so several pressures could precede the same regain.

To establish who controlled the ball when pressure began, we looked for opposing passes, carries, and dribbles with a frame within half a second of the pressure on either side. We required a controlling action at the pressure time, a carry spanning that time, or control actions by the same player on either side. We discarded assignments when another player controlled the ball, the ball had been released or was in flight, or play had stopped.

Once we had identified the carrier in a nearby frame, we matched players in that frame to players in the pressure frame. A pair had to agree on team and goalkeeper status. For complete matching, we allowed a common positional offset of up to 3 meters if the offsets estimated separately for the two teams agreed within 0.5 meters. After alignment, paired players could be at most $0.5+12|\Delta t|$ meters apart, where $\Delta t$ is the time between frames in seconds. We accepted a carrier location only if every feasible complete matching placed the carrier at that point.

When complete matching failed, we tried a partial match using the original coordinates. It had to include players from both teams, at least four players in total, and at least 80\% of the players in the smaller frame. Every matching with the largest feasible number of pairs had to agree on the carrier's location. If usable reference frames disagreed, we excluded the pressure.

We checked carrier identification on events whose carrier was known because the carrier was the recorded actor. We hid the actor and applied the same half-second matching rule, which accepted 20,137 of 713,114 masked targets. Every accepted assignment named the correct player, and 15 (0.07\%) placed that player at the wrong point in the frame. Because a pressure event names only the presser, these known-carrier events provide our reference for assignment error.

To ensure there was room to test positions around the presser, we checked eight equally spaced directions at distances of 1.25 and 2 meters. At each distance at least four directions had to yield a valid position under the pitch, visibility, and player-separation rules in Section~\ref{app:positioning_sampling}. We applied this check before sampling the 64 candidate positions. Of 132,415 recorded pressures across all 380 Premier League fixtures, 20,293 met the carrier, frame, space, and outcome requirements. These involved 476 named pressers.

Of the 20,293 accepted Premier League pressures, 12,843 used a reference event at or before the pressure and 7,450 used one after it. A later reference helped establish who had the ball at the time of pressure. We excluded reference timing and evidence of the subsequent regain from the classifier inputs.

For the high-press experiment, we then required the carrier to be in the possession team's own half, with a visible pressing outfielder within 6.4 meters and a boundary-inclusive pressure score above 0.65 (Section~\ref{app:detector_details}). Applied at the observed positions, these conditions retained 8,607 pressures by 452 players. We kept this set of presses fixed when moving the presser.

\subsection{Regain outcome and model fitting}\label{app:positioning_model}

We started the five-second outcome window at the recorded pressure. A completed pressing-team pass, a carry or shot by that team, or a goalkeeper action that secured the ball (a successful collection, a keeper-sweeper claim, or a won smother) established control directly. A successful recovery, interception, or won tackle required confirmation. Among the next two on-ball events, we required a pressing-team pass, carry, or shot within two seconds and before any opponent contact or stoppage. A pass used for confirmation could be incomplete. A regain beginning within five seconds could therefore be confirmed shortly after the window ended. We counted a favorable restart without prior control as no regain. We excluded observations if the period or event record ended before we could determine the outcome.

We fitted the binary XGBoost classifier using only observed positions from the 20,293 eligible pressures. Each pressure received equal training weight. We calculated the 26 inputs in Table~\ref{tab:features} at the pressure event, used preceding play for context, and constructed the two team shapes independently (Section~\ref{app:model_sensitivities}). We divided matches into five folds without shuffling. Each fit used 250 boosting rounds with XGBoost's histogram method, maximum depth 3, learning rate 0.04, minimum child weight 10, $L_2$ penalty 1, full row and column sampling, and the training-fold regain frequency as its starting probability. We used the resulting binary probabilities directly to score each held-out match, including the alternative positions tested in that match.

Across all eligible pressures, 3,593 ended in controlled regains. Mean predicted regain probability was 17.67\%, compared with 17.71\% observed. Held-out AUC was 0.630, average precision was 0.249 against a base rate of 0.177, and log loss was 0.452 against 0.467 for a constant prediction at the base rate. In the subset of 8,607 high presses used in Fig.~\ref{fig:counterfactual_positions}, 1,473 ended in controlled regains. AUC was 0.625, with a mean predicted regain probability of 17.08\% and an observed rate of 17.11\%.

\subsection{Alternative positions and network reconstruction}\label{app:positioning_sampling}

We sampled positions in a ring around the observed presser so that equal areas of the ring received equal sampling weight. For each eligible high press, we took the first 64 points $(u,v)$ of a two-dimensional scrambled Sobol sequence in the unit square, using a seed derived from the match and event identifiers. We set the displacement to $r=\sqrt{0.25+3.75u}$ meters and its angle to $\theta=2\pi v$ relative to the direction from the presser to the carrier. This gives radii between 0.5 and 2 meters.

A candidate had to lie inside the pitch and the visible area, at least 0.5 meters from every other visible player. We discarded invalid candidates without replacing them. Of 550,848 proposals across 8,607 high presses, 540,886 were valid. Each pressure retained between 32 and 64 alternatives.

At each valid position we moved only the recorded presser and rebuilt the pressure links, passing links, Voronoi cells, and team shapes. We retained the preceding play as observed. For a feature measuring change, we therefore subtracted the same preceding value from the newly calculated current value. We scored all positions for a pressure with the classifier fitted without that match and subtracted the mean alternative prediction from the observed-position prediction (Eq.~\ref{eq:position_response}).

Moving a presser changes the pressure on the carrier and, because the preceding play was held fixed, also the change in carrier pressure since the preceding observation. To see how much the scores depended on this second input, we rescored the same 8,607 presses and 540,886 alternatives with the same fitted models, holding the change in carrier pressure at its observed value while recalculating current pressure and space at each alternative. Mean position quality fell from 0.391 to 0.067 percentage points (95\% fixture-bootstrap interval $[0.038,\,0.096]$). When we also held the changes in carrier space and forward-receiver space at their observed values, it was 0.028 percentage points ($[0.002,\,0.054]$). Across the 181 ranked players, Spearman's correlation with the original scores was 0.79 with the pressure change held fixed and 0.74 with all three changes held fixed, and five and three of the original top ten, respectively, remained in the top ten. The correlation between club means and season points fell from 0.58 to 0.25 and 0.27. The estimated advantage therefore depended substantially on updating the features that describe how the press had developed.

\subsection{Aggregation and uncertainty}\label{app:positioning_summaries}

To calculate club and player means, we gave each eligible high press equal weight. We assigned each press to the pressing club and combined a player's presses across clubs using the provider's player identifier. The player ranking required at least 15 eligible high presses and 1,350 Premier League minutes. We summed unrounded provider minutes across fixtures and clubs before applying the minutes threshold. The 181 players who met both conditions entered Fig.~\ref{fig:counterfactual_positions}c. Club means included every eligible press. For a 95\% interval around an individual player or club mean, we resampled its contributing fixtures with replacement 2,000 times, keeping all eligible high presses from each sampled fixture.

For the association between mean position quality and season points across the 20 clubs, we drew 380 whole fixtures with replacement in each of 10,000 draws. Within each draw, we recomputed the club means together and correlated them with the clubs' actual final points totals. The observed Spearman correlation was 0.576, with a 95\% percentile interval of $[-0.035,\,0.643]$ (Fig.~\ref{fig:counterfactual_positions}b). To keep the number of fixtures per club fixed, we also resampled 38 complete official matchweeks with replacement in each of 10,000 draws. Each club played once per matchweek, giving it 38 fixtures in every draw. The corresponding interval was $[0.015,\,0.624]$.

\subsection{Sampling and temporal checks}\label{app:positioning_checks}

To check whether 64 proposals were sufficient to estimate the local average, we recalculated each pressure's score using the valid positions among the first 32 proposals. Both calculations used the same observed position, classifier, and ordering of proposals. Across all 8,607 high presses, the median absolute difference was 0.052 percentage points and the 95th percentile was 0.218 percentage points.

We estimated each player's mean separately for fixtures played before January 1, 2026 and for fixtures played on or after that date, requiring at least five eligible high presses across at least three matches in each half. Both halves used the same match-held-out scores from classifiers fitted across the whole season. Among the 197 players who qualified, Spearman's correlation between the two rankings was 0.067 (approximate 95\% interval $[-0.07,\,0.20]$). With half a season of presses per player, two rankings would agree only partly even if each player's positioning were stable. To gauge how much, we reassigned each player's matches to the two halves at random, keeping the number of that player's matches in each half, so that any lasting difference between players appeared equally in both halves. Over 1,000 such reassignments, the median correlation was 0.13 (central 95\% range 0.03 to 0.22). When we instead shuffled presses among players within each half, which removes any lasting difference between players, the median was $-0.01$ ($-0.14$ to $0.13$). The observed correlation fell in the lower part of the first range and the upper part of the second. Separating lasting positional habits from the particular situations players faced will therefore require more presses per player than half a season provides.

\subsection{Applying the regain model to the 2026 men's World Cup}\label{app:positioning_world_cup}

The 2026 men's World Cup provided a separate test of the regain model trained on Premier League pressures. We applied the same carrier-identification, pressure-eligibility, and outcome rules to the 103 of 104 fixtures with both event and 360 data. We identified 1,360 eligible high presses and obtained 84,928 valid alternatives using the sampling procedure in Section~\ref{app:positioning_sampling}. At each observed and alternative position, we averaged the five Premier League classifiers' regain probabilities. We then calculated position quality within each pressure and averaged those scores within each player. No World Cup data were used to fit the classifiers.

We evaluated the predictions at the observed positions against the recorded outcomes. There were 219 controlled regains within five seconds among 1,360 high presses. The AUC was 0.641 (95\% percentile interval $[0.602,\,0.678]$, resampling 103 whole fixtures with replacement in each of 10,000 draws). Mean position quality across the high presses was 0.37 percentage points, compared with 0.39 in the Premier League. For the player ranking, we counted eligible pressures in the 103 fixtures with frames and tournament minutes in all 104 fixtures. Sixty-three players had at least five eligible high presses and at least 180 minutes.

\section{High-press position quality by player}\label{sec:si-high-press-position-quality}

Table~\ref{tab:si-high-press-position-quality} lists every player who logged Premier League minutes in 2025/26 and ranks the 452 who had at least one eligible high press by their mean position quality (Eq.~\ref{eq:position_response}). Figure~\ref{fig:counterfactual_positions}c shows the ten highest means among the 181 players with at least 15 eligible high presses and at least 1,350 minutes, who appear in bold. Players with few presses fill both ends of the table: each of the 26 players ranked above Jayden Bogle, first in Fig.~\ref{fig:counterfactual_positions}c, had eight or fewer eligible high presses, and each of the 37 ranked below the lowest-ranked player who qualified for the figure had 13 or fewer.

{
\footnotesize
\setlength{\tabcolsep}{4pt}
\setlength{\LTcapwidth}{\linewidth}

\begin{longtable}{@{}r >{\raggedright\arraybackslash}p{0.30\linewidth} >{\raggedright\arraybackslash}p{0.24\linewidth} r r r@{}}
\caption{\textbf{High-press position quality in the 2025/26 Premier League.} The table lists all 537 players who logged league minutes. Scores are mean observed-minus-alternative differences in predicted five-second regain probability, in percentage points per high press. Ranks use unrounded scores across all players with a score. Exact ties share a rank. Bold rows identify players with at least 15 eligible high presses and 1,350 league minutes. A dash indicates that the player had no eligible high presses.}\label{tab:si-high-press-position-quality}\\
\toprule
\textbf{Rank} & \textbf{Player} & \textbf{Club(s)} & \textbf{Minutes} & \shortstack[r]{\textbf{High}\\\textbf{presses}} & \shortstack[r]{\textbf{Position}\\\textbf{quality (pp)}} \\
\midrule
\endfirsthead
\multicolumn{6}{@{}l}{\textit{Table \thetable\ continued}} \\
\toprule
\textbf{Rank} & \textbf{Player} & \textbf{Club(s)} & \textbf{Minutes} & \shortstack[r]{\textbf{High}\\\textbf{presses}} & \shortstack[r]{\textbf{Position}\\\textbf{quality (pp)}} \\
\midrule
\endhead
\midrule
\multicolumn{6}{r@{}}{\textit{Continued on the next page}} \\
\endfoot
\bottomrule
\endlastfoot
1 & Virgil van Dijk & Liverpool & 3,848.7 & 1 & +4.638 \\
2 & Arthur Masuaku & Sunderland & 141.5 & 1 & +4.373 \\
3 & Christian N\o{}rgaard & Arsenal & 145.0 & 1 & +4.150 \\
4 & Matt Doherty & Wolves & 852.6 & 2 & +3.803 \\
5 & Nathan Collins & Brentford & 3,332.6 & 6 & +3.789 \\
6 & Dane Scarlett & Tottenham Hotspur & 20.3 & 1 & +3.415 \\
7 & Mateo Kova\v{c}i\'{c} & Manchester City & 170.2 & 3 & +3.234 \\
8 & Sa\v{s}a Kalajd\v{z}i\'{c} & Wolves & 21.3 & 1 & +2.732 \\
9 & Jorge Cuenca & Fulham & 1,047.3 & 5 & +2.729 \\
10 & Emmanuel Agbadou & Wolves & 1,228.3 & 2 & +2.686 \\
11 & Adam Smith & AFC Bournemouth & 1,210.7 & 8 & +2.556 \\
12 & Luke O'Nien & Sunderland & 679.1 & 1 & +2.427 \\
13 & Myles Lewis-Skelly & Arsenal & 848.0 & 3 & +2.352 \\
14 & Wataru Endo & Liverpool & 222.4 & 1 & +2.304 \\
15 & Arnaud Kalimuendo & Nottingham Forest & 144.4 & 5 & +2.153 \\
16 & Alex T\'{o}th & AFC Bournemouth & 198.1 & 4 & +2.141 \\
17 & Victor Lindel\"{o}f & Aston Villa & 1,063.3 & 1 & +2.123 \\
18 & Morato & Nottingham Forest & 1,525.7 & 3 & +2.029 \\
19 & David M\o{}ller Wolfe & Wolves & 1,185.1 & 7 & +1.994 \\
20 & Facundo Buonanotte & Chelsea; Leeds United & 96.6 & 4 & +1.921 \\
21 & Andy Irving & West Ham & 206.5 & 5 & +1.761 \\
22 & Max Alleyne & Manchester City & 146.9 & 1 & +1.758 \\
23 & Ibrahima Konat\'{e} & Liverpool & 3,431.4 & 7 & +1.751 \\
24 & Matthijs de Ligt & Manchester United & 1,308.0 & 8 & +1.748 \\
25 & Noussair Mazraoui & Manchester United & 1,112.6 & 8 & +1.729 \\
26 & William Saliba & Arsenal & 2,934.3 & 6 & +1.727 \\
\textbf{27} & \textbf{Jayden Bogle} & \textbf{Leeds United} & \textbf{3,043.0} & \textbf{22} & \textbf{+1.667} \\
\textbf{28} & \textbf{Adrien Truffert} & \textbf{AFC Bournemouth} & \textbf{3,772.4} & \textbf{18} & \textbf{+1.532} \\
29 & Tosin Adarabioyo & Chelsea & 935.8 & 3 & +1.522 \\
30 & Bafod\'{e} Diakit\'{e} & AFC Bournemouth & 1,448.7 & 4 & +1.493 \\
31 & Cristian Romero & Tottenham Hotspur & 2,057.2 & 11 & +1.492 \\
\textbf{32} & \textbf{Rodri} & \textbf{Manchester City} & \textbf{1,676.0} & \textbf{16} & \textbf{+1.482} \\
\textbf{33} & \textbf{Luke Shaw} & \textbf{Manchester United} & \textbf{3,513.4} & \textbf{19} & \textbf{+1.448} \\
34 & Chadi Riad & Crystal Palace & 619.1 & 1 & +1.414 \\
\textbf{35} & \textbf{Mats Wieffer} & \textbf{Brighton} & \textbf{2,108.8} & \textbf{20} & \textbf{+1.409} \\
36 & Guido Rodr\'{\i}guez & West Ham & 204.8 & 5 & +1.398 \\
37 & Joe Gomez & Liverpool & 723.4 & 4 & +1.394 \\
38 & Marcos Senesi & AFC Bournemouth & 3,672.8 & 8 & +1.382 \\
39 & Andy Robertson & Liverpool & 1,325.1 & 2 & +1.369 \\
40 & Konstantinos Mavropanos & West Ham & 2,820.7 & 3 & +1.360 \\
\textbf{41} & \textbf{Lewis Hall} & \textbf{Newcastle United} & \textbf{2,472.0} & \textbf{20} & \textbf{+1.355} \\
42 & Destiny Udogie & Tottenham Hotspur & 1,483.9 & 7 & +1.349 \\
43 & Souza & Tottenham Hotspur & 157.8 & 1 & +1.339 \\
\textbf{44} & \textbf{Vitalii Mykolenko} & \textbf{Everton} & \textbf{3,262.4} & \textbf{20} & \textbf{+1.331} \\
45 & Ashley Barnes & Burnley & 99.7 & 2 & +1.318 \\
46 & Igor Julio & West Ham & 132.2 & 2 & +1.315 \\
47 & James Tarkowski & Everton & 3,678.5 & 4 & +1.313 \\
48 & Jo\v{s}ko Gvardiol & Manchester City & 1,514.3 & 8 & +1.302 \\
49 & Pau Torres & Aston Villa & 1,884.6 & 4 & +1.247 \\
50 & Merlin R\"{o}hl & Everton & 767.8 & 13 & +1.244 \\
51 & Ladislav Krej\v{c}\'{\i} & Wolves & 2,594.5 & 6 & +1.206 \\
52 & Harrison Reed & Fulham & 134.8 & 4 & +1.183 \\
53 & Wilfried Gnonto & Leeds United & 686.8 & 11 & +1.167 \\
54 & Fer L\'{o}pez & Wolves & 287.9 & 8 & +1.152 \\
\textbf{55} & \textbf{Wilson Isidor} & \textbf{Sunderland} & \textbf{1,356.2} & \textbf{19} & \textbf{+1.147} \\
56 & Trevoh Chalobah & Chelsea & 3,145.4 & 5 & +1.145 \\
57 & Tammy Abraham & Aston Villa & 356.8 & 5 & +1.142 \\
\textbf{58} & \textbf{Sander Berge} & \textbf{Fulham} & \textbf{3,210.9} & \textbf{42} & \textbf{+1.123} \\
59 & Marc Guiu & Chelsea; Sunderland & 333.3 & 6 & +1.120 \\
60 & Jean-Clair Todibo & West Ham & 2,020.8 & 4 & +1.118 \\
61 & Joe Willock & Newcastle United & 1,140.4 & 19 & +1.114 \\
\textbf{62} & \textbf{Jurri\"{e}n Timber} & \textbf{Arsenal} & \textbf{2,737.8} & \textbf{21} & \textbf{+1.095} \\
63 & Tyrone Mings & Aston Villa & 1,471.8 & 1 & +1.077 \\
64 & Frank Onyeka & Brentford & 126.6 & 2 & +1.074 \\
65 & Matt O'Riley & Brighton & 233.2 & 9 & +1.071 \\
\textbf{66} & \textbf{Andr\'{e}} & \textbf{Wolves} & \textbf{3,066.3} & \textbf{47} & \textbf{+1.069} \\
\textbf{67} & \textbf{Carlos Baleba} & \textbf{Brighton} & \textbf{1,787.5} & \textbf{32} & \textbf{+1.063} \\
68 & Kenny Tete & Fulham & 1,971.1 & 9 & +1.061 \\
69 & Joachim Andersen & Fulham & 3,193.4 & 13 & +1.048 \\
\textbf{70} & \textbf{Declan Rice} & \textbf{Arsenal} & \textbf{3,440.2} & \textbf{16} & \textbf{+1.026} \\
71 & Sepp van den Berg & Brentford & 3,112.6 & 3 & +1.022 \\
72 & Daniel James & Leeds United & 655.0 & 9 & +1.019 \\
73 & Axel Disasi & West Ham & 1,375.1 & 5 & +1.012 \\
74 & Jan Paul van Hecke & Brighton & 3,578.6 & 14 & +1.007 \\
75 & Pascal Struijk & Leeds United & 3,265.8 & 11 & +1.003 \\
76 & James Hill & AFC Bournemouth & 2,375.0 & 7 & +0.993 \\
77 & Lamare Bogarde & Aston Villa & 1,213.1 & 21 & +0.992 \\
78 & Freddie Potts & West Ham & 1,296.4 & 13 & +0.988 \\
79 & Zeki Amdouni & Burnley & 102.0 & 1 & +0.987 \\
80 & Malick Thiaw & Newcastle United & 3,365.3 & 9 & +0.986 \\
81 & Reinildo Mandava & Sunderland & 2,199.7 & 11 & +0.983 \\
\textbf{82} & \textbf{Curtis Jones} & \textbf{Liverpool} & \textbf{2,206.4} & \textbf{30} & \textbf{+0.970} \\
83 & Jamaldeen Jimoh-Aloba & Aston Villa & 15.9 & 1 & +0.968 \\
84 & Michael Keane & Everton & 2,875.8 & 8 & +0.955 \\
85 & R\'{u}ben Dias & Manchester City & 2,389.8 & 6 & +0.944 \\
86 & Lutsharel Geertruida & Sunderland & 1,924.6 & 10 & +0.936 \\
\textbf{87} & \textbf{Youri Tielemans} & \textbf{Aston Villa} & \textbf{2,028.1} & \textbf{31} & \textbf{+0.932} \\
\textbf{88} & \textbf{Michael Kayode} & \textbf{Brentford} & \textbf{3,624.1} & \textbf{29} & \textbf{+0.925} \\
89 & Ben Davies & Tottenham Hotspur & 147.2 & 2 & +0.921 \\
90 & Jo\"{e}l Veltman & Brighton & 1,226.6 & 12 & +0.912 \\
\textbf{91} & \textbf{Hannibal Mejbri} & \textbf{Burnley} & \textbf{1,378.0} & \textbf{27} & \textbf{+0.912} \\
92 & Maxence Lacroix & Crystal Palace & 3,419.5 & 10 & +0.902 \\
\textbf{93} & \textbf{Ryan Sessegnon} & \textbf{Fulham} & \textbf{2,044.3} & \textbf{17} & \textbf{+0.897} \\
94 & Ola Aina & Nottingham Forest & 1,745.1 & 9 & +0.893 \\
\textbf{95} & \textbf{Neco Williams} & \textbf{Nottingham Forest} & \textbf{3,530.1} & \textbf{25} & \textbf{+0.862} \\
96 & Jaydee Canvot & Crystal Palace & 1,507.5 & 8 & +0.851 \\
\textbf{97} & \textbf{Mart\'{\i}n Zubimendi} & \textbf{Arsenal} & \textbf{3,305.9} & \textbf{40} & \textbf{+0.844} \\
\textbf{98} & \textbf{Josh Cullen} & \textbf{Burnley} & \textbf{1,667.0} & \textbf{18} & \textbf{+0.824} \\
99 & Gabriel Magalh\~{a}es & Arsenal & 3,089.2 & 5 & +0.814 \\
\textbf{100} & \textbf{Nordi Mukiele} & \textbf{Sunderland} & \textbf{3,084.5} & \textbf{18} & \textbf{+0.794} \\
101 & Manuel Ugarte & Manchester United & 1,024.2 & 18 & +0.793 \\
102 & Max Kilman & West Ham & 1,758.5 & 4 & +0.787 \\
103 & Tom\'{a}\v{s} Sou\v{c}ek & West Ham & 2,467.5 & 10 & +0.786 \\
\textbf{104} & \textbf{Rodrigo Bentancur} & \textbf{Tottenham Hotspur} & \textbf{2,075.3} & \textbf{21} & \textbf{+0.783} \\
105 & Jo\"{e}l Piroe & Leeds United & 360.6 & 3 & +0.781 \\
106 & Jacob Bruun Larsen & Burnley & 1,184.5 & 23 & +0.772 \\
\textbf{107} & \textbf{Igor Jesus} & \textbf{Nottingham Forest} & \textbf{2,498.3} & \textbf{51} & \textbf{+0.761} \\
\textbf{108} & \textbf{Harvey Barnes} & \textbf{Newcastle United} & \textbf{2,229.2} & \textbf{24} & \textbf{+0.761} \\
\textbf{109} & \textbf{Tyler Adams} & \textbf{AFC Bournemouth} & \textbf{1,951.6} & \textbf{36} & \textbf{+0.743} \\
110 & Charalampos Kostoulas & Brighton & 497.7 & 16 & +0.739 \\
111 & Dennis Cirkin & Sunderland & 369.3 & 8 & +0.732 \\
\textbf{112} & \textbf{Jarrod Bowen} & \textbf{West Ham} & \textbf{3,750.5} & \textbf{36} & \textbf{+0.718} \\
113 & Chris Richards & Crystal Palace & 3,159.6 & 10 & +0.718 \\
114 & Murillo & Nottingham Forest & 2,336.5 & 8 & +0.713 \\
115 & Gabriel Martinelli & Arsenal & 1,258.0 & 16 & +0.711 \\
116 & John Stones & Manchester City & 503.8 & 2 & +0.705 \\
117 & Taiwo Awoniyi & Nottingham Forest & 604.7 & 8 & +0.702 \\
\textbf{118} & \textbf{Will Hughes} & \textbf{Crystal Palace} & \textbf{1,761.7} & \textbf{31} & \textbf{+0.697} \\
\textbf{119} & \textbf{Ferdi Kad\i{}o\u{g}lu} & \textbf{Brighton} & \textbf{3,480.6} & \textbf{30} & \textbf{+0.697} \\
120 & Pablo Felipe & West Ham & 901.1 & 34 & +0.693 \\
\textbf{121} & \textbf{Marc Cucurella} & \textbf{Chelsea} & \textbf{3,044.3} & \textbf{21} & \textbf{+0.691} \\
122 & Rodrigo Muniz & Fulham & 1,107.4 & 11 & +0.691 \\
\textbf{123} & \textbf{Idrissa Gueye} & \textbf{Everton} & \textbf{2,279.9} & \textbf{40} & \textbf{+0.689} \\
\textbf{124} & \textbf{Riccardo Calafiori} & \textbf{Arsenal} & \textbf{1,876.8} & \textbf{17} & \textbf{+0.687} \\
125 & Gabriel Jesus & Arsenal & 502.9 & 11 & +0.677 \\
126 & Sven Botman & Newcastle United & 2,075.2 & 6 & +0.673 \\
127 & Alexander Isak & Liverpool & 769.4 & 10 & +0.673 \\
\textbf{128} & \textbf{Alex Scott} & \textbf{AFC Bournemouth} & \textbf{3,143.0} & \textbf{48} & \textbf{+0.670} \\
\textbf{129} & \textbf{Florentino Lu\'{\i}s} & \textbf{Burnley} & \textbf{2,308.9} & \textbf{30} & \textbf{+0.669} \\
\textbf{130} & \textbf{Enzo Fern\'{a}ndez} & \textbf{Chelsea} & \textbf{3,485.1} & \textbf{50} & \textbf{+0.666} \\
\textbf{131} & \textbf{Taty Castellanos} & \textbf{West Ham} & \textbf{1,471.0} & \textbf{29} & \textbf{+0.666} \\
\textbf{132} & \textbf{Jordan Henderson} & \textbf{Brentford} & \textbf{2,155.5} & \textbf{28} & \textbf{+0.660} \\
\textbf{133} & \textbf{Granit Xhaka} & \textbf{Sunderland} & \textbf{3,243.6} & \textbf{34} & \textbf{+0.652} \\
\textbf{134} & \textbf{Phil Foden} & \textbf{Manchester City} & \textbf{2,313.3} & \textbf{55} & \textbf{+0.640} \\
135 & Nathaniel Clyne & Crystal Palace & 659.8 & 11 & +0.633 \\
\textbf{136} & \textbf{Nico O'Reilly} & \textbf{Manchester City} & \textbf{2,920.2} & \textbf{40} & \textbf{+0.626} \\
\textbf{137} & \textbf{Iliman Ndiaye} & \textbf{Everton} & \textbf{3,041.6} & \textbf{36} & \textbf{+0.626} \\
138 & Donyell Malen & Aston Villa & 755.9 & 11 & +0.626 \\
\textbf{139} & \textbf{Mikkel Damsgaard} & \textbf{Brentford} & \textbf{2,275.1} & \textbf{50} & \textbf{+0.620} \\
\textbf{140} & \textbf{Jean-Ricner Bellegarde} & \textbf{Wolves} & \textbf{1,385.0} & \textbf{20} & \textbf{+0.620} \\
141 & Jeremie Frimpong & Liverpool & 1,176.9 & 17 & +0.620 \\
\textbf{142} & \textbf{James Justin} & \textbf{Leeds United} & \textbf{2,157.8} & \textbf{20} & \textbf{+0.619} \\
\textbf{143} & \textbf{Ilia Gruev} & \textbf{Leeds United} & \textbf{1,480.0} & \textbf{23} & \textbf{+0.615} \\
\textbf{144} & \textbf{Patrick Dorgu} & \textbf{Manchester United} & \textbf{1,616.6} & \textbf{19} & \textbf{+0.615} \\
\textbf{145} & \textbf{Maxim De Cuyper} & \textbf{Brighton} & \textbf{1,597.0} & \textbf{22} & \textbf{+0.612} \\
\textbf{146} & \textbf{Leandro Trossard} & \textbf{Arsenal} & \textbf{2,191.4} & \textbf{26} & \textbf{+0.600} \\
\textbf{147} & \textbf{Mohamed Salah} & \textbf{Liverpool} & \textbf{2,403.5} & \textbf{26} & \textbf{+0.597} \\
148 & Alysson & Aston Villa & 69.4 & 1 & +0.593 \\
\textbf{149} & \textbf{Djed Spence} & \textbf{Tottenham Hotspur} & \textbf{2,336.2} & \textbf{22} & \textbf{+0.590} \\
\textbf{150} & \textbf{Matheus Nunes} & \textbf{Manchester City} & \textbf{3,168.3} & \textbf{24} & \textbf{+0.589} \\
\textbf{151} & \textbf{Lucas Pires} & \textbf{Burnley} & \textbf{1,591.4} & \textbf{26} & \textbf{+0.588} \\
152 & Stefanos Tzimas & Brighton & 196.4 & 10 & +0.584 \\
\textbf{153} & \textbf{Dominik Szoboszlai} & \textbf{Liverpool} & \textbf{3,628.6} & \textbf{63} & \textbf{+0.582} \\
\textbf{154} & \textbf{Sandro Tonali} & \textbf{Newcastle United} & \textbf{2,812.8} & \textbf{60} & \textbf{+0.577} \\
155 & Ben Doak & AFC Bournemouth & 177.0 & 2 & +0.572 \\
\textbf{156} & \textbf{Kobbie Mainoo} & \textbf{Manchester United} & \textbf{1,896.7} & \textbf{20} & \textbf{+0.570} \\
\textbf{157} & \textbf{Marcus Tavernier} & \textbf{AFC Bournemouth} & \textbf{2,981.4} & \textbf{53} & \textbf{+0.568} \\
158 & Chris Wood & Nottingham Forest & 962.3 & 18 & +0.553 \\
\textbf{159} & \textbf{Jack Grealish} & \textbf{Everton} & \textbf{1,774.4} & \textbf{21} & \textbf{+0.545} \\
160 & Santiago Bueno & Wolves & 2,840.1 & 6 & +0.545 \\
161 & Samuel Chukwueze & Fulham & 1,204.2 & 14 & +0.544 \\
162 & Nayef Aguerd & West Ham & 196.9 & 1 & +0.544 \\
\textbf{163} & \textbf{Yankuba Minteh} & \textbf{Brighton} & \textbf{2,634.2} & \textbf{49} & \textbf{+0.542} \\
164 & Amadou Onana & Aston Villa & 1,945.9 & 12 & +0.536 \\
165 & Aaron Hickey & Brentford & 800.1 & 11 & +0.535 \\
\textbf{166} & \textbf{Alexis Mac Allister} & \textbf{Liverpool} & \textbf{2,980.6} & \textbf{58} & \textbf{+0.535} \\
\textbf{167} & \textbf{Lesley Ugochukwu} & \textbf{Burnley} & \textbf{2,543.2} & \textbf{40} & \textbf{+0.530} \\
\textbf{168} & \textbf{Tyrick Mitchell} & \textbf{Crystal Palace} & \textbf{3,608.9} & \textbf{52} & \textbf{+0.527} \\
169 & Ryan Christie & AFC Bournemouth & 1,115.5 & 17 & +0.524 \\
\textbf{170} & \textbf{Pascal Gro\ss{}} & \textbf{Brighton} & \textbf{1,818.9} & \textbf{29} & \textbf{+0.521} \\
171 & Wesley Fofana & Chelsea & 1,908.2 & 4 & +0.520 \\
\textbf{172} & \textbf{Noah Sadiki} & \textbf{Sunderland} & \textbf{3,222.6} & \textbf{46} & \textbf{+0.517} \\
173 & Lukas Nmecha & Leeds United & 1,274.3 & 24 & +0.511 \\
\textbf{174} & \textbf{Nico Gonz\'{a}lez} & \textbf{Manchester City} & \textbf{1,750.2} & \textbf{26} & \textbf{+0.509} \\
175 & Sean Longstaff & Leeds United & 1,181.0 & 37 & +0.508 \\
\textbf{176} & \textbf{Lucas Paquet\'{a}} & \textbf{West Ham} & \textbf{1,647.3} & \textbf{30} & \textbf{+0.498} \\
177 & F\'{a}bio Carvalho & Brentford & 140.5 & 5 & +0.498 \\
\textbf{178} & \textbf{Bruno Fernandes} & \textbf{Manchester United} & \textbf{3,397.9} & \textbf{81} & \textbf{+0.497} \\
\textbf{179} & \textbf{Jacob Ramsey} & \textbf{Newcastle United} & \textbf{1,680.5} & \textbf{31} & \textbf{+0.494} \\
\textbf{180} & \textbf{Gabriel Gudmundsson} & \textbf{Leeds United} & \textbf{2,892.0} & \textbf{26} & \textbf{+0.493} \\
181 & Marcus Edwards & Burnley & 1,185.0 & 14 & +0.493 \\
\textbf{182} & \textbf{Mois\'{e}s Caicedo} & \textbf{Chelsea} & \textbf{3,127.5} & \textbf{35} & \textbf{+0.490} \\
\textbf{183} & \textbf{Reece James} & \textbf{Chelsea} & \textbf{2,206.6} & \textbf{16} & \textbf{+0.486} \\
\textbf{184} & \textbf{Anthony Gordon} & \textbf{Newcastle United} & \textbf{1,951.4} & \textbf{49} & \textbf{+0.479} \\
185 & Jadon Sancho & Aston Villa & 979.9 & 4 & +0.478 \\
\textbf{186} & \textbf{J\'{e}r\'{e}my Doku} & \textbf{Manchester City} & \textbf{1,953.4} & \textbf{24} & \textbf{+0.475} \\
\textbf{187} & \textbf{Jacob Murphy} & \textbf{Newcastle United} & \textbf{1,820.6} & \textbf{43} & \textbf{+0.471} \\
\textbf{188} & \textbf{Kevin Schade} & \textbf{Brentford} & \textbf{3,000.5} & \textbf{32} & \textbf{+0.469} \\
\textbf{189} & \textbf{Yehor Yarmoliuk} & \textbf{Brentford} & \textbf{2,915.8} & \textbf{65} & \textbf{+0.465} \\
190 & Jhon Arias & Wolves & 1,253.8 & 29 & +0.464 \\
191 & Quilindschy Hartman & Burnley & 2,006.8 & 11 & +0.464 \\
192 & Jefferson Lerma & Crystal Palace & 1,981.1 & 10 & +0.463 \\
\textbf{193} & \textbf{Keane Lewis-Potter} & \textbf{Brentford} & \textbf{2,209.4} & \textbf{24} & \textbf{+0.462} \\
\textbf{194} & \textbf{Mateus Fernandes} & \textbf{West Ham} & \textbf{3,299.2} & \textbf{41} & \textbf{+0.461} \\
195 & Piero Hincapi\'{e} & Arsenal & 2,000.3 & 8 & +0.456 \\
\textbf{196} & \textbf{Ibrahim Sangar\'{e}} & \textbf{Nottingham Forest} & \textbf{2,279.5} & \textbf{39} & \textbf{+0.456} \\
197 & James Milner & Brighton & 906.0 & 14 & +0.453 \\
\textbf{198} & \textbf{Tijjani Reijnders} & \textbf{Manchester City} & \textbf{1,789.9} & \textbf{37} & \textbf{+0.450} \\
\textbf{199} & \textbf{Bruno Guimar\~{a}es} & \textbf{Newcastle United} & \textbf{2,758.6} & \textbf{59} & \textbf{+0.448} \\
200 & Christantus Uche & Crystal Palace & 259.0 & 9 & +0.448 \\
\textbf{201} & \textbf{John McGinn} & \textbf{Aston Villa} & \textbf{2,303.3} & \textbf{16} & \textbf{+0.448} \\
\textbf{202} & \textbf{Adam Wharton} & \textbf{Crystal Palace} & \textbf{2,800.3} & \textbf{40} & \textbf{+0.444} \\
\textbf{203} & \textbf{Matty Cash} & \textbf{Aston Villa} & \textbf{3,312.7} & \textbf{21} & \textbf{+0.442} \\
\textbf{204} & \textbf{Vitaly Janelt} & \textbf{Brentford} & \textbf{1,647.5} & \textbf{28} & \textbf{+0.439} \\
\textbf{205} & \textbf{Malo Gusto} & \textbf{Chelsea} & \textbf{2,518.8} & \textbf{22} & \textbf{+0.438} \\
206 & Andrey Santos & Chelsea & 1,417.0 & 14 & +0.429 \\
\textbf{207} & \textbf{Martin \O{}degaard} & \textbf{Arsenal} & \textbf{1,513.5} & \textbf{21} & \textbf{+0.423} \\
208 & Mohamadou Kant\'{e} & West Ham & 126.8 & 3 & +0.408 \\
\textbf{209} & \textbf{J\o{}rgen Strand Larsen} & \textbf{Crystal Palace; Wolves} & \textbf{2,544.7} & \textbf{50} & \textbf{+0.406} \\
210 & Lisandro Mart\'{\i}nez & Manchester United & 1,367.8 & 4 & +0.405 \\
211 & Sebastiaan Bornauw & Leeds United & 480.6 & 2 & +0.405 \\
\textbf{212} & \textbf{Ollie Watkins} & \textbf{Aston Villa} & \textbf{3,077.7} & \textbf{37} & \textbf{+0.397} \\
\textbf{213} & \textbf{Enzo Le F\'{e}e} & \textbf{Sunderland} & \textbf{3,256.4} & \textbf{58} & \textbf{+0.388} \\
\textbf{214} & \textbf{Anthony Elanga} & \textbf{Newcastle United} & \textbf{1,535.7} & \textbf{22} & \textbf{+0.381} \\
215 & Timothy Castagne & Fulham & 2,092.8 & 7 & +0.380 \\
\textbf{216} & \textbf{Georginio Rutter} & \textbf{Brighton} & \textbf{1,955.0} & \textbf{37} & \textbf{+0.379} \\
217 & Rom\'{e}o Lavia & Chelsea & 458.9 & 5 & +0.376 \\
\textbf{218} & \textbf{Jackson Tchatchoua} & \textbf{Wolves} & \textbf{2,073.0} & \textbf{15} & \textbf{+0.373} \\
\textbf{219} & \textbf{Richarlison} & \textbf{Tottenham Hotspur} & \textbf{2,213.0} & \textbf{47} & \textbf{+0.366} \\
\textbf{220} & \textbf{Viktor Gy\"{o}keres} & \textbf{Arsenal} & \textbf{2,487.5} & \textbf{33} & \textbf{+0.362} \\
\textbf{221} & \textbf{Bernardo Silva} & \textbf{Manchester City} & \textbf{3,106.2} & \textbf{73} & \textbf{+0.361} \\
222 & Kevin Danso & Tottenham Hotspur & 1,722.0 & 2 & +0.360 \\
223 & Callum Marshall & West Ham & 71.9 & 4 & +0.357 \\
\textbf{224} & \textbf{Yasin Ayari} & \textbf{Brighton} & \textbf{2,166.0} & \textbf{38} & \textbf{+0.357} \\
\textbf{225} & \textbf{Callum Hudson-Odoi} & \textbf{Nottingham Forest} & \textbf{1,997.0} & \textbf{22} & \textbf{+0.354} \\
226 & Conor Bradley & Liverpool & 1,030.0 & 10 & +0.351 \\
\textbf{227} & \textbf{Sa\v{s}a Luki\'{c}} & \textbf{Fulham} & \textbf{1,916.0} & \textbf{44} & \textbf{+0.349} \\
228 & Olivier Boscagli & Brighton & 900.0 & 1 & +0.344 \\
229 & Savinho & Manchester City & 979.2 & 9 & +0.336 \\
230 & Eddie Nketiah & Crystal Palace & 510.6 & 6 & +0.333 \\
231 & Lewis Dunk & Brighton & 3,172.5 & 7 & +0.332 \\
\textbf{232} & \textbf{Erling Haaland} & \textbf{Manchester City} & \textbf{3,255.8} & \textbf{44} & \textbf{+0.331} \\
\textbf{233} & \textbf{Anton Stach} & \textbf{Leeds United} & \textbf{2,598.5} & \textbf{44} & \textbf{+0.330} \\
\textbf{234} & \textbf{Eberechi Eze} & \textbf{Arsenal; Cr.~Palace} & \textbf{2,104.6} & \textbf{37} & \textbf{+0.329} \\
\textbf{235} & \textbf{Casemiro} & \textbf{Manchester United} & \textbf{2,767.2} & \textbf{34} & \textbf{+0.324} \\
\textbf{236} & \textbf{Ra\'{u}l Jim\'{e}nez} & \textbf{Fulham} & \textbf{2,411.6} & \textbf{41} & \textbf{+0.323} \\
\textbf{237} & \textbf{Micky van de Ven} & \textbf{Tottenham Hotspur} & \textbf{3,417.5} & \textbf{21} & \textbf{+0.323} \\
238 & Ryan Yates & Nottingham Forest & 754.4 & 11 & +0.321 \\
\textbf{239} & \textbf{Rayan Cherki} & \textbf{Manchester City} & \textbf{1,986.2} & \textbf{20} & \textbf{+0.320} \\
240 & James McAtee & Nottingham Forest & 402.0 & 12 & +0.317 \\
\textbf{241} & \textbf{Matheus Cunha} & \textbf{Manchester United} & \textbf{2,725.8} & \textbf{43} & \textbf{+0.317} \\
242 & Simon Adingra & Sunderland & 717.4 & 16 & +0.315 \\
243 & Maxime Est\`{e}ve & Burnley & 3,244.0 & 3 & +0.312 \\
\textbf{244} & \textbf{Antoine Semenyo} & \textbf{Bournemo.; Man.~City} & \textbf{3,531.1} & \textbf{56} & \textbf{+0.305} \\
245 & Ezri Konsa & Aston Villa & 3,367.2 & 5 & +0.294 \\
\textbf{246} & \textbf{Amad Diallo} & \textbf{Manchester United} & \textbf{2,593.2} & \textbf{27} & \textbf{+0.292} \\
247 & Tyler Dibling & Everton & 419.9 & 8 & +0.292 \\
248 & Evann Guessand & Aston Villa; Cr.~Palace & 1,063.0 & 10 & +0.291 \\
\textbf{249} & \textbf{Pedro Porro} & \textbf{Tottenham Hotspur} & \textbf{3,132.8} & \textbf{15} & \textbf{+0.290} \\
\textbf{250} & \textbf{Elliot Anderson} & \textbf{Nottingham Forest} & \textbf{3,657.2} & \textbf{72} & \textbf{+0.289} \\
251 & Dan Burn & Newcastle United & 2,492.9 & 13 & +0.289 \\
\textbf{252} & \textbf{Alex Iwobi} & \textbf{Fulham} & \textbf{2,620.9} & \textbf{20} & \textbf{+0.286} \\
253 & Radu Dr\u{a}gu\textcommabelow{s}in & Tottenham Hotspur & 607.5 & 1 & +0.284 \\
254 & Dwight McNeil & Everton & 1,271.4 & 21 & +0.278 \\
\textbf{255} & \textbf{Jo\~{a}o Pedro} & \textbf{Chelsea} & \textbf{2,933.5} & \textbf{43} & \textbf{+0.278} \\
\textbf{256} & \textbf{Emile Smith Rowe} & \textbf{Fulham} & \textbf{2,109.6} & \textbf{54} & \textbf{+0.273} \\
257 & Nikola Milenkovi\'{c} & Nottingham Forest & 3,730.3 & 6 & +0.266 \\
\textbf{258} & \textbf{Trai Hume} & \textbf{Sunderland} & \textbf{3,380.3} & \textbf{36} & \textbf{+0.261} \\
\textbf{259} & \textbf{Florian Wirtz} & \textbf{Liverpool} & \textbf{2,608.6} & \textbf{48} & \textbf{+0.257} \\
260 & El Hadji Malick Diouf & West Ham & 2,969.4 & 11 & +0.256 \\
261 & Jorrel Hato & Chelsea & 1,308.8 & 10 & +0.252 \\
\textbf{262} & \textbf{Mateus Man\'{e}} & \textbf{Wolves} & \textbf{2,021.9} & \textbf{52} & \textbf{+0.251} \\
\textbf{263} & \textbf{Alejandro Garnacho} & \textbf{Chelsea} & \textbf{1,423.1} & \textbf{23} & \textbf{+0.251} \\
\textbf{264} & \textbf{Danny Welbeck} & \textbf{Brighton} & \textbf{2,479.2} & \textbf{56} & \textbf{+0.251} \\
\textbf{265} & \textbf{Jean-Philippe Mateta} & \textbf{Crystal Palace} & \textbf{2,428.8} & \textbf{34} & \textbf{+0.247} \\
266 & Nicol\'{a}s Dom\'{\i}nguez & Nottingham Forest & 1,304.5 & 34 & +0.239 \\
267 & Tom Cairney & Fulham & 915.9 & 8 & +0.237 \\
268 & Adam Armstrong & Wolves & 1,251.0 & 27 & +0.235 \\
\textbf{269} & \textbf{Nick Woltemade} & \textbf{Newcastle United} & \textbf{2,100.0} & \textbf{74} & \textbf{+0.234} \\
270 & Romelle Donovan & Brentford & 69.6 & 1 & +0.232 \\
\textbf{271} & \textbf{David Brooks} & \textbf{AFC Bournemouth} & \textbf{1,376.6} & \textbf{20} & \textbf{+0.232} \\
272 & Marshall Munetsi & Wolves & 973.2 & 27 & +0.231 \\
273 & Yoane Wissa & Newcastle United & 658.2 & 10 & +0.230 \\
\textbf{274} & \textbf{Aaron Wan-Bissaka} & \textbf{West Ham} & \textbf{2,318.8} & \textbf{19} & \textbf{+0.229} \\
\textbf{275} & \textbf{Igor Thiago} & \textbf{Brentford} & \textbf{3,646.3} & \textbf{84} & \textbf{+0.224} \\
\textbf{276} & \textbf{Dominic Calvert-Lewin} & \textbf{Leeds United} & \textbf{3,001.8} & \textbf{50} & \textbf{+0.224} \\
\textbf{277} & \textbf{Noah Okafor} & \textbf{Leeds United} & \textbf{1,702.9} & \textbf{27} & \textbf{+0.222} \\
278 & Carlos Alcaraz & Everton & 800.6 & 20 & +0.221 \\
\textbf{279} & \textbf{Kyle Walker} & \textbf{Burnley} & \textbf{3,441.9} & \textbf{39} & \textbf{+0.219} \\
\textbf{280} & \textbf{Hugo Ekitik\'{e}} & \textbf{Liverpool} & \textbf{1,990.2} & \textbf{29} & \textbf{+0.215} \\
\textbf{281} & \textbf{Thierno Barry} & \textbf{Everton} & \textbf{2,116.3} & \textbf{49} & \textbf{+0.215} \\
\textbf{282} & \textbf{Morgan Gibbs-White} & \textbf{Nottingham Forest} & \textbf{3,356.0} & \textbf{61} & \textbf{+0.207} \\
\textbf{283} & \textbf{Mathys Tel} & \textbf{Tottenham Hotspur} & \textbf{1,554.0} & \textbf{22} & \textbf{+0.206} \\
284 & Liam Delap & Chelsea & 1,295.9 & 23 & +0.202 \\
\textbf{285} & \textbf{Tim Iroegbunam} & \textbf{Everton} & \textbf{1,660.5} & \textbf{22} & \textbf{+0.202} \\
286 & Jake O'Brien & Everton & 3,422.6 & 6 & +0.202 \\
\textbf{287} & \textbf{Lyle Foster} & \textbf{Burnley} & \textbf{1,525.3} & \textbf{24} & \textbf{+0.190} \\
288 & Noni Madueke & Arsenal & 1,335.6 & 21 & +0.187 \\
\textbf{289} & \textbf{Dango Ouattara} & \textbf{Brentford} & \textbf{2,517.0} & \textbf{44} & \textbf{+0.186} \\
290 & Oscar Bobb & Fulham; Manchester City & 1,182.3 & 28 & +0.185 \\
\textbf{291} & \textbf{Hugo Bueno} & \textbf{Wolves} & \textbf{2,638.1} & \textbf{23} & \textbf{+0.181} \\
292 & Dominic Solanke & Tottenham Hotspur & 1,125.3 & 21 & +0.180 \\
\textbf{293} & \textbf{Bryan Mbeumo} & \textbf{Manchester United} & \textbf{2,838.3} & \textbf{45} & \textbf{+0.176} \\
294 & Tom Edozie & Wolves & 25.6 & 2 & +0.163 \\
\textbf{295} & \textbf{Jo\~{a}o Palhinha} & \textbf{Tottenham Hotspur} & \textbf{2,486.2} & \textbf{40} & \textbf{+0.157} \\
\textbf{296} & \textbf{Daichi Kamada} & \textbf{Crystal Palace} & \textbf{2,080.9} & \textbf{28} & \textbf{+0.150} \\
\textbf{297} & \textbf{Hwang Hee-chan} & \textbf{Wolves} & \textbf{1,574.9} & \textbf{33} & \textbf{+0.149} \\
\textbf{298} & \textbf{Mathias Jensen} & \textbf{Brentford} & \textbf{2,490.2} & \textbf{41} & \textbf{+0.145} \\
\textbf{299} & \textbf{Yeremy Pino} & \textbf{Crystal Palace} & \textbf{2,255.0} & \textbf{45} & \textbf{+0.145} \\
300 & Leon Bailey & Aston Villa & 547.2 & 6 & +0.134 \\
301 & Est\^{e}v\~{a}o & Chelsea & 952.0 & 11 & +0.127 \\
\textbf{302} & \textbf{Pedro Neto} & \textbf{Chelsea} & \textbf{2,874.9} & \textbf{22} & \textbf{+0.127} \\
303 & Justin Devenny & Crystal Palace & 921.5 & 18 & +0.123 \\
304 & Conor Gallagher & Tottenham Hotspur & 1,323.7 & 18 & +0.122 \\
\textbf{305} & \textbf{Diego G\'{o}mez} & \textbf{Brighton} & \textbf{2,284.6} & \textbf{97} & \textbf{+0.120} \\
\textbf{306} & \textbf{Joelinton} & \textbf{Newcastle United} & \textbf{2,186.5} & \textbf{39} & \textbf{+0.106} \\
\textbf{307} & \textbf{Crysencio Summerville} & \textbf{West Ham} & \textbf{2,705.1} & \textbf{18} & \textbf{+0.102} \\
308 & Bertrand Traor\'{e} & Sunderland & 702.1 & 6 & +0.097 \\
\textbf{309} & \textbf{Josh King} & \textbf{Fulham} & \textbf{1,461.1} & \textbf{31} & \textbf{+0.096} \\
\textbf{310} & \textbf{Morgan Rogers} & \textbf{Aston Villa} & \textbf{3,623.3} & \textbf{55} & \textbf{+0.095} \\
\textbf{311} & \textbf{Randal Kolo Muani} & \textbf{Tottenham Hotspur} & \textbf{1,851.8} & \textbf{25} & \textbf{+0.094} \\
\textbf{312} & \textbf{Jo\~{a}o Gomes} & \textbf{Wolves} & \textbf{3,110.6} & \textbf{51} & \textbf{+0.088} \\
\textbf{313} & \textbf{Benjamin \v{S}e\v{s}ko} & \textbf{Manchester United} & \textbf{1,863.8} & \textbf{35} & \textbf{+0.077} \\
314 & Romaine Mundle & Sunderland & 494.7 & 19 & +0.077 \\
\textbf{315} & \textbf{Ethan Ampadu} & \textbf{Leeds United} & \textbf{3,462.2} & \textbf{21} & \textbf{+0.075} \\
316 & Harrison Armstrong & Everton & 588.7 & 19 & +0.075 \\
\textbf{317} & \textbf{Beto} & \textbf{Everton} & \textbf{1,744.5} & \textbf{36} & \textbf{+0.072} \\
318 & James Ward-Prowse & Burnley; West Ham & 1,210.4 & 19 & +0.067 \\
319 & Justin Kluivert & AFC Bournemouth & 1,071.2 & 21 & +0.063 \\
320 & Toti Gomes & Wolves & 1,587.6 & 1 & +0.063 \\
\textbf{321} & \textbf{Cole Palmer} & \textbf{Chelsea} & \textbf{2,174.4} & \textbf{15} & \textbf{+0.061} \\
\textbf{322} & \textbf{Harry Wilson} & \textbf{Fulham} & \textbf{2,890.9} & \textbf{26} & \textbf{+0.061} \\
\textbf{323} & \textbf{Kiernan Dewsbury-Hall} & \textbf{Everton} & \textbf{2,851.3} & \textbf{88} & \textbf{+0.059} \\
324 & Cristhian Mosquera & Arsenal & 1,132.4 & 6 & +0.059 \\
\textbf{325} & \textbf{Brenden Aaronson} & \textbf{Leeds United} & \textbf{2,638.1} & \textbf{66} & \textbf{+0.057} \\
\textbf{326} & \textbf{Kaoru Mitoma} & \textbf{Brighton} & \textbf{1,890.7} & \textbf{30} & \textbf{+0.055} \\
\textbf{327} & \textbf{Kieran Trippier} & \textbf{Newcastle United} & \textbf{1,764.3} & \textbf{15} & \textbf{+0.055} \\
328 & Nilson Angulo & Sunderland & 447.1 & 6 & +0.046 \\
329 & Mikel Merino & Arsenal & 1,186.4 & 16 & +0.044 \\
\textbf{330} & \textbf{Josh Laurent} & \textbf{Burnley} & \textbf{1,975.2} & \textbf{20} & \textbf{+0.044} \\
\textbf{331} & \textbf{Jaidon Anthony} & \textbf{Burnley} & \textbf{2,978.6} & \textbf{61} & \textbf{+0.042} \\
332 & Jack Fletcher & Manchester United & 125.6 & 7 & +0.041 \\
333 & Archie Gray & Tottenham Hotspur & 1,671.0 & 8 & +0.040 \\
\textbf{334} & \textbf{Xavi Simons} & \textbf{Tottenham Hotspur} & \textbf{1,971.6} & \textbf{25} & \textbf{+0.039} \\
\textbf{335} & \textbf{Jack Hinshelwood} & \textbf{Brighton} & \textbf{1,896.9} & \textbf{42} & \textbf{+0.038} \\
336 & Dilane Bakwa & Nottingham Forest & 623.5 & 11 & +0.032 \\
\textbf{337} & \textbf{Evanilson} & \textbf{AFC Bournemouth} & \textbf{2,952.0} & \textbf{34} & \textbf{+0.030} \\
\textbf{338} & \textbf{\'{A}lex Jim\'{e}nez} & \textbf{AFC Bournemouth} & \textbf{2,535.4} & \textbf{17} & \textbf{+0.029} \\
339 & Ezra Mayers & West Ham & 152.7 & 1 & +0.028 \\
\textbf{340} & \textbf{Brian Brobbey} & \textbf{Sunderland} & \textbf{2,111.8} & \textbf{24} & \textbf{+0.027} \\
\textbf{341} & \textbf{Omari Hutchinson} & \textbf{Nottingham Forest} & \textbf{1,868.9} & \textbf{31} & \textbf{+0.026} \\
342 & Enes \"{U}nal & AFC Bournemouth & 374.8 & 8 & +0.021 \\
343 & Nicol\`{o} Savona & Nottingham Forest & 1,140.4 & 10 & +0.018 \\
344 & Hjalmar Ekdal & Burnley & 1,702.8 & 5 & +0.018 \\
345 & Jaka Bijol & Leeds United & 2,135.4 & 4 & +0.017 \\
346 & Soungoutou Magassa & West Ham & 1,096.6 & 16 & +0.015 \\
\textbf{347} & \textbf{Eli Junior Kroupi} & \textbf{AFC Bournemouth} & \textbf{1,823.9} & \textbf{35} & \textbf{+0.014} \\
348 & Rayan & AFC Bournemouth & 1,186.4 & 13 & +0.006 \\
349 & Oleksandr Zinchenko & Nottingham Forest & 380.9 & 1 & +0.002 \\
350 & Boubacar Kamara & Aston Villa & 1,539.1 & 14 & +0.001 \\
\textbf{351} & \textbf{Cody Gakpo} & \textbf{Liverpool} & \textbf{2,989.9} & \textbf{25} & \textbf{-0.002} \\
\textbf{352} & \textbf{James Garner} & \textbf{Everton} & \textbf{3,771.5} & \textbf{45} & \textbf{-0.002} \\
\textbf{353} & \textbf{Loum Tchaouna} & \textbf{Burnley} & \textbf{1,444.0} & \textbf{34} & \textbf{-0.005} \\
\textbf{354} & \textbf{Tolu Arokodare} & \textbf{Wolves} & \textbf{1,642.6} & \textbf{30} & \textbf{-0.008} \\
355 & Fabian Sch\"{a}r & Newcastle United & 1,243.5 & 5 & -0.008 \\
356 & Pedro Lima & Wolves & 311.1 & 2 & -0.016 \\
357 & Lewis Cook & AFC Bournemouth & 1,015.3 & 11 & -0.017 \\
358 & Antonee Robinson & Fulham & 1,661.1 & 12 & -0.020 \\
359 & Ethan Nwaneri & Arsenal & 214.7 & 4 & -0.020 \\
\textbf{360} & \textbf{Daniel Mu\~{n}oz} & \textbf{Crystal Palace} & \textbf{2,651.7} & \textbf{32} & \textbf{-0.021} \\
361 & Milos Kerkez & Liverpool & 2,536.6 & 8 & -0.022 \\
362 & Joshua Zirkzee & Manchester United & 774.1 & 15 & -0.029 \\
\textbf{363} & \textbf{Zian Flemming} & \textbf{Burnley} & \textbf{1,942.7} & \textbf{35} & \textbf{-0.030} \\
\textbf{364} & \textbf{Mohammed Kudus} & \textbf{Tottenham Hotspur} & \textbf{1,662.6} & \textbf{43} & \textbf{-0.031} \\
\textbf{365} & \textbf{Isma\"{\i}la Sarr} & \textbf{Crystal Palace} & \textbf{2,405.4} & \textbf{32} & \textbf{-0.031} \\
366 & Omar Marmoush & Manchester City & 810.9 & 14 & -0.033 \\
367 & Eliezer Mayenda & Sunderland & 862.0 & 13 & -0.036 \\
\textbf{368} & \textbf{Habib Diarra} & \textbf{Sunderland} & \textbf{1,562.3} & \textbf{33} & \textbf{-0.037} \\
369 & Bashir Humphreys & Burnley & 1,748.7 & 6 & -0.045 \\
370 & Amine Adli & AFC Bournemouth & 1,271.9 & 19 & -0.045 \\
371 & Ethan Pinnock & Brentford & 341.0 & 2 & -0.047 \\
372 & Federico Chiesa & Liverpool & 524.3 & 7 & -0.047 \\
\textbf{373} & \textbf{Emiliano Buend\'{\i}a} & \textbf{Aston Villa} & \textbf{1,944.2} & \textbf{35} & \textbf{-0.057} \\
374 & Mason Mount & Manchester United & 1,148.0 & 23 & -0.068 \\
375 & Lucas Bergvall & Tottenham Hotspur & 1,087.4 & 24 & -0.069 \\
376 & Dan Ndoye & Nottingham Forest & 1,292.0 & 23 & -0.070 \\
377 & Rico Lewis & Manchester City & 482.3 & 3 & -0.073 \\
\textbf{378} & \textbf{Brennan Johnson} & \textbf{Cr.~Palace; Tottenham} & \textbf{1,915.9} & \textbf{34} & \textbf{-0.079} \\
379 & Douglas Luiz & A.~Villa; Nottingham For. & 1,019.2 & 15 & -0.081 \\
380 & Joe Rodon & Leeds United & 3,298.0 & 6 & -0.088 \\
381 & Ki-Jana Hoever & Wolves & 460.8 & 5 & -0.090 \\
382 & Rayan A\"{\i}t-Nouri & Manchester City & 1,094.1 & 7 & -0.093 \\
383 & Tawanda Chirewa & Wolves & 41.2 & 1 & -0.105 \\
\textbf{384} & \textbf{Bukayo Saka} & \textbf{Arsenal} & \textbf{2,474.6} & \textbf{36} & \textbf{-0.108} \\
385 & Ao Tanaka & Leeds United & 1,465.9 & 11 & -0.108 \\
386 & Harvey Elliott & Aston Villa; Liverpool & 148.5 & 2 & -0.109 \\
387 & Rico Henry & Brentford & 1,503.0 & 11 & -0.112 \\
388 & Oliver Sonne & Burnley & 207.7 & 4 & -0.114 \\
389 & William Osula & Newcastle United & 974.1 & 20 & -0.120 \\
390 & Shea Lacey & Manchester United & 57.9 & 1 & -0.127 \\
391 & Jaydon Banel & Burnley & 12.9 & 2 & -0.133 \\
392 & Kevin & Fulham & 1,194.2 & 13 & -0.134 \\
393 & Ryan Gravenberch & Liverpool & 3,341.3 & 14 & -0.142 \\
394 & Chris Rigg & Sunderland & 848.6 & 9 & -0.143 \\
395 & Jamie Gittens & Chelsea & 574.7 & 10 & -0.148 \\
396 & Brajan Gruda & Brighton & 823.1 & 18 & -0.149 \\
397 & Tyrique George & Chelsea; Everton & 464.8 & 11 & -0.173 \\
398 & Ollie Scarles & West Ham & 754.9 & 4 & -0.181 \\
\textbf{399} & \textbf{Lewis Miley} & \textbf{Newcastle United} & \textbf{1,706.9} & \textbf{21} & \textbf{-0.183} \\
400 & Issa Diop & Fulham & 926.4 & 2 & -0.192 \\
401 & Josh Acheampong & Chelsea & 764.7 & 3 & -0.194 \\
402 & Armando Broja & Burnley & 1,003.0 & 18 & -0.198 \\
403 & Calvin Bassey & Fulham & 2,840.3 & 10 & -0.198 \\
404 & Adama Traor\'{e} & Fulham; West Ham & 551.4 & 4 & -0.212 \\
\textbf{405} & \textbf{Chemsdine Talbi} & \textbf{Sunderland} & \textbf{1,710.9} & \textbf{23} & \textbf{-0.214} \\
406 & Tino Livramento & Newcastle United & 1,504.1 & 13 & -0.232 \\
407 & Niclas F\"{u}llkrug & West Ham & 452.2 & 6 & -0.235 \\
408 & Lorenzo Lucca & Nottingham Forest & 155.0 & 2 & -0.238 \\
409 & Ross Barkley & Aston Villa & 956.4 & 12 & -0.260 \\
410 & Kyle Walker-Peters & West Ham & 1,550.4 & 13 & -0.262 \\
411 & Nathan Patterson & Everton & 279.1 & 3 & -0.265 \\
412 & Yerson Mosquera & Wolves & 2,356.2 & 14 & -0.280 \\
413 & Pape Matar Sarr & Tottenham Hotspur & 1,612.4 & 12 & -0.309 \\
414 & Kristoffer Ajer & Brentford & 2,026.2 & 9 & -0.318 \\
\textbf{415} & \textbf{Callum Wilson} & \textbf{West Ham} & \textbf{1,452.3} & \textbf{24} & \textbf{-0.337} \\
416 & Axel Tuanzebe & Burnley & 1,341.3 & 3 & -0.350 \\
417 & Jack Harrison & Leeds United & 323.2 & 5 & -0.366 \\
418 & Emiliano Mart\'{\i}nez & Aston Villa & 3,150.6 & 1 & -0.391 \\
419 & Marc Gu\'{e}hi & Cr.~Palace; Man.~City & 3,499.8 & 7 & -0.407 \\
420 & Wilson Odobert & Tottenham Hotspur & 1,121.3 & 13 & -0.409 \\
421 & Angel Gomes & Wolves & 515.5 & 9 & -0.422 \\
422 & Bradley Burrows & Aston Villa & 23.6 & 3 & -0.437 \\
423 & Yves Bissouma & Tottenham Hotspur & 685.2 & 9 & -0.450 \\
424 & Diogo Dalot & Manchester United & 2,895.8 & 12 & -0.468 \\
425 & Kai Havertz & Arsenal & 646.0 & 6 & -0.472 \\
426 & Julio Soler & AFC Bournemouth & 49.1 & 4 & -0.474 \\
427 & Beno\^{\i}t Badiashile & Chelsea & 529.7 & 1 & -0.499 \\
428 & Ben White & Arsenal & 771.0 & 5 & -0.505 \\
429 & Ian Maatsen & Aston Villa & 1,791.1 & 10 & -0.517 \\
430 & Ayden Heaven & Manchester United & 1,037.6 & 3 & -0.522 \\
431 & Leny Yoro & Manchester United & 1,960.8 & 6 & -0.553 \\
432 & Jonah Kusi-Asare & Fulham & 94.8 & 3 & -0.555 \\
433 & Divine Mukasa & Manchester City & 36.5 & 1 & -0.581 \\
434 & Lucas Digne & Aston Villa & 2,060.0 & 4 & -0.603 \\
435 & Joel Drakes-Thomas & Crystal Palace & 25.1 & 1 & -0.611 \\
436 & Luis Guilherme & West Ham & 149.4 & 3 & -0.728 \\
437 & Rodrigo Gomes & Wolves & 1,098.5 & 12 & -0.736 \\
438 & George Earthy & West Ham & 11.7 & 1 & -0.753 \\
439 & Rio Ngumoha & Liverpool & 682.7 & 8 & -0.774 \\
440 & Luca Netz & Nottingham Forest & 417.2 & 5 & -0.842 \\
441 & Junai Byfield & Tottenham Hotspur & 9.5 & 1 & -0.884 \\
442 & Mike Tr\'{e}sor & Burnley & 75.9 & 1 & -0.931 \\
443 & Rio Cardines & Crystal Palace & 101.2 & 1 & -1.118 \\
444 & Max Dowman & Arsenal & 189.4 & 2 & -1.207 \\
445 & Borna Sosa & Crystal Palace & 142.8 & 1 & -1.233 \\
446 & Tommy Watson & Brighton & 89.0 & 2 & -1.247 \\
447 & Joe Worrall & Burnley & 592.4 & 1 & -1.430 \\
448 & Reiss Nelson & Brentford & 186.5 & 2 & -1.649 \\
449 & Harry Maguire & Manchester United & 1,832.0 & 1 & -2.086 \\
450 & Omar Alderete & Sunderland & 3,110.1 & 2 & -2.388 \\
451 & Jocelin Ta Bi & Sunderland & 38.3 & 1 & -2.418 \\
452 & Dan Ballard & Sunderland & 2,409.0 & 1 & -3.257 \\
- & Aaron Ramsdale & Newcastle United & 1,147.8 & 0 & - \\
- & Abdukodir Khusanov & Manchester City & 1,587.5 & 0 & - \\
- & Alex Murphy & Newcastle United & 22.4 & 0 & - \\
- & Alisson & Liverpool & 2,634.1 & 0 & - \\
- & Alphonse Areola & West Ham & 2,015.1 & 0 & - \\
- & Altay Bay\i{}nd\i{}r & Manchester United & 609.2 & 0 & - \\
- & Andr\'{e}s Garc\'{\i}a & Aston Villa & 118.9 & 0 & - \\
- & Angus Gunn & Nottingham Forest & 48.3 & 0 & - \\
- & Antoni Milambo & Brentford & 49.6 & 0 & - \\
- & Anton\'{\i}n Kinsk\'{y} & Tottenham Hotspur & 725.7 & 0 & - \\
- & Bart Verbruggen & Brighton & 3,811.9 & 0 & - \\
- & Ben Winterburn & AFC Bournemouth & 13.0 & 0 & - \\
- & Bendito Mantato & Manchester United & 22.4 & 0 & - \\
- & Bernd Leno & Fulham & 3,812.4 & 0 & - \\
- & Callum Olusesi & Tottenham Hotspur & 21.3 & 0 & - \\
- & Caoimh\'{\i}n Kelleher & Brentford & 3,719.0 & 0 & - \\
- & Dan Bentley & Wolves & 301.7 & 0 & - \\
- & Dan Neil & Sunderland & 25.8 & 0 & - \\
- & D\'{a}rio Essugo & Chelsea & 49.0 & 0 & - \\
- & David Raya & Arsenal & 3,728.6 & 0 & - \\
- & Dean Henderson & Crystal Palace & 3,712.6 & 0 & - \\
- & Diego Coppola & Brighton & 218.7 & 0 & - \\
- & Emil Krafth & Newcastle United & 51.0 & 0 & - \\
- & Filip J\"{o}rgensen & Chelsea & 439.2 & 0 & - \\
- & Freddie Woodman & Liverpool & 246.7 & 0 & - \\
- & George Hemmings & Aston Villa & 38.1 & 0 & - \\
- & Gianluigi Donnarumma & Manchester City & 3,386.0 & 0 & - \\
- & Giorgi Mamardashvili & Liverpool & 968.0 & 0 & - \\
- & Guglielmo Vicario & Tottenham Hotspur & 3,129.3 & 0 & - \\
- & Hamed Traor\`{e} & AFC Bournemouth & 22.4 & 0 & - \\
- & Harry Howell & Brighton & 110.3 & 0 & - \\
- & H\'{a}kon Valdimarsson & Brentford & 107.5 & 0 & - \\
- & Jair Cunha & Nottingham Forest & 594.5 & 0 & - \\
- & Jamaal Lascelles & Newcastle United & 33.4 & 0 & - \\
- & James Maddison & Tottenham Hotspur & 69.2 & 0 & - \\
- & James Rowswell & Tottenham Hotspur & 2.0 & 0 & - \\
- & James Trafford & Manchester City & 412.6 & 0 & - \\
- & Jarrad Branthwaite & Everton & 742.2 & 0 & - \\
- & Jenson Seelt & Sunderland & 148.5 & 0 & - \\
- & Jerome Abbey & Wolves & 22.9 & 0 & - \\
- & Jesse Derry & Chelsea & 54.0 & 0 & - \\
- & Joe Knight & Brighton & 1.8 & 0 & - \\
- & John Victor & Nottingham Forest & 473.2 & 0 & - \\
- & Jordan Pickford & Everton & 3,782.2 & 0 & - \\
- & Josh Dasilva & Brentford & 21.5 & 0 & - \\
- & Jos\'{e} S\'{a} & Wolves & 2,303.7 & 0 & - \\
- & Jota Silva & Nottingham Forest & 16.4 & 0 & - \\
- & Karl Darlow & Leeds United & 2,203.0 & 0 & - \\
- & Kayden Rodney & Crystal Palace & 8.8 & 0 & - \\
- & Kaye Furo & Brentford & 8.5 & 0 & - \\
- & Kepa Arrizabalaga & Arsenal & 101.2 & 0 & - \\
- & Levi Colwill & Chelsea & 259.7 & 0 & - \\
- & Lucas Perri & Leeds United & 1,619.9 & 0 & - \\
- & Mads Hermansen & West Ham & 1,788.2 & 0 & - \\
- & Mamadou Sarr & Chelsea & 118.5 & 0 & - \\
- & Marco Bizot & Aston Villa & 649.6 & 0 & - \\
- & Martin D\'{u}bravka & Burnley & 3,507.2 & 0 & - \\
- & Matz Sels & Nottingham Forest & 2,970.9 & 0 & - \\
- & Max Wei\ss{} & Burnley & 308.5 & 0 & - \\
- & Melker Ellborg & Sunderland & 308.8 & 0 & - \\
- & Myles Peart-Harris & Brentford & 3.1 & 0 & - \\
- & Nathan Ak\'{e} & Manchester City & 807.7 & 0 & - \\
- & Nehemiah Oriola & Brighton & 1.8 & 0 & - \\
- & Nick Pope & Newcastle United & 2,727.6 & 0 & - \\
- & Odsonne \'{E}douard & Crystal Palace & 7.5 & 0 & - \\
- & Patrick Roberts & Sunderland & 37.6 & 0 & - \\
- & Robert S\'{a}nchez & Chelsea & 3,407.8 & 0 & - \\
- & Robin Roefs & Sunderland & 3,517.5 & 0 & - \\
- & Romain Esse & Crystal Palace & 65.1 & 0 & - \\
- & Sam Byram & Leeds United & 23.1 & 0 & - \\
- & Sam Johnstone & Wolves & 1,207.2 & 0 & - \\
- & Sean Neave & Newcastle United & 21.3 & 0 & - \\
- & Senne Lammens & Manchester United & 3,198.5 & 0 & - \\
- & Shumaira Mheuka & Chelsea & 8.9 & 0 & - \\
- & Solly March & Brighton & 64.4 & 0 & - \\
- & Stefan Ortega & Nottingham Forest & 297.3 & 0 & - \\
- & S\'{e}amus Coleman & Everton & 40.1 & 0 & - \\
- & Trey Nyoni & Liverpool & 53.9 & 0 & - \\
- & Tyler Fletcher & Manchester United & 26.4 & 0 & - \\
- & Tyler Fredricson & Manchester United & 9.3 & 0 & - \\
- & Tyrell Malacia & Manchester United & 37.9 & 0 & - \\
- & Veljko Milosavljevi\'{c} & AFC Bournemouth & 412.4 & 0 & - \\
- & Walter Ben\'{\i}tez & Crystal Palace & 101.8 & 0 & - \\
- & Zach Abbott & Nottingham Forest & 179.6 & 0 & - \\
- & \DJ{}or\dj{}e Petrovi\'{c} & AFC Bournemouth & 3,823.6 & 0 & - \\
\end{longtable}
}

\printbibliography[title={Supplemental References}]
\end{refsection}

\end{document}